\documentclass[11pt]{article}
\usepackage{jheppub}
\pdfoutput=1
\usepackage[utf8]{inputenc}
\usepackage[T1]{fontenc}
\usepackage{amsmath}
\usepackage{amssymb}
\usepackage[bb=boondox]{mathalpha}
\usepackage{bm}
\usepackage[Symbolsmallscale]{upgreek}
\usepackage{physics}
\usepackage{mathtools}
\usepackage{array}
\usepackage{tikz}
\usepackage{standalone}
\usepackage{graphicx}
\usepackage{enumitem}
\usepackage[hang, bottom]{footmisc}
\usepackage{xcolor}
\usepackage{colortbl}
\usepackage{multirow}
\usepackage{hhline}
\usepackage{braket}
\usepackage[noabbrev]{cleveref}

\usetikzlibrary{positioning}

\allowdisplaybreaks

\newcommand{\ii}{\operatorname{i}}
\newcommand{\col}{\operatorname{col}}

\newcommand{\arctanh}{\operatorname{arctanh}}

\newcommand{\ad}{\operatorname{ad}}

\newcommand{\ee}{\operatorname{e}}

\renewcommand{\cosh}{\operatorname{ch}}
\renewcommand{\sinh}{\operatorname{sh}}
\renewcommand{\hat}{\widehat}
\renewcommand{\tilde}{\widetilde}

\newcommand{\Intk}{\int \dd^d{\vec{k}} \:}

\title{Symmetries and the Hilbert Space of Large \emph{N} Extended States}

\author[a,b]{Antal Jevicki,}
\author[a,b]{Xianlong Liu,}
\author[a,b]{Junjie Zheng}

 \affiliation[a]{Department of Physics, Brown University, \\
                182 Hope Street, Providence, RI 02912, USA}
 \affiliation[b]{Brown Theoretical Physics Center, Brown University, \\ 
                340 Brook Street, Providence, RI 02912, USA}

\emailAdd{antal\_jevicki@brown.edu}
\emailAdd{xianlong\_liu@brown.edu}
\emailAdd{junjie\_zheng@brown.edu}

\dedicated{Dedicated to the memory of Jean-Loup Gervais.}

\abstract{We discuss the large $N$ expansion in backgrounds of extended states with focus on implementation of Goldstone symmetries and the construction of the associated Hilbert space. The formulation is given in the general framework of collective field theory. Case of translational symmetry is described first, as the basic example. The large $N$ thermofield represents the main topics, 
with the emergent dynamics of Left-Right bulk fields and collective symmetry coordinates. These give the basis for a $1/N$ expansion.
}

\begin{document}

\maketitle

\section{Introduction}

Large $N$ represents the basic non-perturbative scheme for implementing AdS/CFT and more generally gauge/gravity duality. 
With $1/N$ representing Newton's constant $G$, the main confirmations of the duality are most easily seen in comparison of correlation functions. Beyond that, of even greater interests are non-perturbative extended states, especially at finite temperature. In particular the thermofield double (TFD) state conjectured to be dual to a two sided black hole \cite{Maldacena:2001kr} is of central interest. So is the structure of the emergent Hilbert space and of the associated $1/N$ expansion. The construction of such TFD state and the development of $1/N$ expansion represents a challenge due to its non-perturbative nature. For vector-type models it was given in \cite{Jevicki:2015sla,Jevicki:2021ddf} where a dynamical symmetry structure (with Goldstone mode properties) was described. Recently investigations \cite{Leutheusser:2021qhd,Leutheusser:2021frk} of the Hilbert space were seen in the gravitational setting with discussions of observables (and of propagation) inside the horizon. In particular, a general discussion of a possible large $N$ Hilbert space structure of the thermal state was presented in \cite{Witten:2021unn,Chandrasekaran:2022eqq} on general grounds.

\paragraph{}
In this work we pursue the construction of the Hilbert space and associate large $N$ expansion in constructive terms. The general framework is collective field theory, which for vector \cite{Das:2003vw,deMelloKoch:2018ivk} (and simple matrix) models is solvable. More generally (for multi-matrix models) it can be studied through numerical optimizations~\cite{Koch:2021yeb}. Therefore for a large class of large $N$ theories large $N$ fluctuations and the emergent structure can be explicitly studied. In the present work we concentrate on the implementation of symmetry structures at Large $N$. Much like in the case of extended (soliton) states in ordinary QFT \cite{Gervais:1975pa} central for the understanding of perturbation expansion is the implementation of relevant (Goldstone) symmetries. These are characterized by large $O(N)$ leading fluctuations (with associated zero modes). The implementation of these symmetries is done through collective coordinates with an associated Hilbert space. We discuss this in the Large $N$ context and describe a formulation of the full nonlinear theory. 

\paragraph{}
Translations are discussed first (providing the basic example) while the main focus is on the thermofield case. Here the question of what the appropriate gauge and Goldstone symmetries already require some analysis. Regarding gauging (of U($N$) and O($N$) groups) we follow the proposal introduced in \cite{Jevicki:2015sla} which gave arguments for diagonal gauging of the doubled large $N$ Hilbert space. It was argued that this is required in the high temperature phase while the low temperature phase can be described by direct product gauging. Fluctuations and the zero mode structure of the large $N$ TFD was furthermore studied in \cite{Jevicki:2015sla,Jevicki:2021ddf} and a relation with a dynamical symmetry was identified. Both of the above features will play a central role in developing the structure of the Hilbert space and of the $1/N$ expansion. For concreteness we follow the vector-model where explicit analysis is viable. The emergent structure and the method of implementing symmetries (through collective coordinates) is however general. 

\paragraph{}
The content of the paper is as follows: \Cref{Sec:Large_N_Hamiltonian} gives a summary of Hamiltonian at large $N$. The case of translations is then discussed first in \Cref{Sec:Translations}, giving a basic example. \Cref{sec:TFD_state_at_large_N} concentrates on the TFD state, and the corresponding degeneracy and symmetry structure. \Cref{sec:colcoord} describes the associated Hilbert space and the general implementation of Goldstone symmetry in thermal case.

\section{The Large \texorpdfstring{$N$}{N} Hamiltonian}
\label{Sec:Large_N_Hamiltonian}

Large $N$ QFT in the canonical formalism, can be completely described through the dynamics of the collective singlet fields $\Phi$. This dynamics is governed by the collective Hamiltonian, generally of the form
\begin{equation}
    H_{\col} = \frac{1}{2} \Pi \, \Omega \, \Pi + V_{\col}[\Phi] \, , 
\end{equation}
with $\Pi = - \ii \delta / \delta \Phi$ the canonical conjugate of $\Phi$. For example, in the large $N$ multi-matrix quantum mechanics case, the collective fields are loop variables. Consider two-matrix systems, let $C = M_1^{n_1} M_2^{n_2} M_1^{n_1^{\prime}} M_2^{n_2^{\prime}} \dots$ denote a word built from the alphabet of two matrices $M_1$ and $M_2$, then the loop variable is defined as  $\Phi(C) = \operatorname{tr}(C)$. The basic terms of the Collective Hamiltonian involve loop joining and splitting operations. In particular,
\begin{equation}
    \Omega(C_1, C_2) = \sum_{C} j(C_1, C_2; C) \Phi(C)
\end{equation}
represents all possible ways of the joining two loop variables $\Phi(C_1)$ and $\Phi(C_2)$ into one loop variable. $j(C_1, C_2; C)$ counts the number of ways that $C$ can be obtained by joining $C_1$ and $C_2$. Similarly,
\begin{equation}
     \omega(C) = \sum_{(C_1, C_2)} p(C; C_1, C_2) \Phi(C_1) \Phi(C_2)
\end{equation}
represents a splitting operation. The integers $p(C; C_1, C_2)$ count the numbers of ways that $C_1$ and $C_2$ can be obtained by splitting $C$. In this case we have the collective potential as
\begin{equation}
    V_{\col}[\Phi] = \frac{1}{8} \omega \Omega^{-1} \omega + V[\Phi] \, ,
\end{equation}
with $V[\Phi]$ the original potential represented by $\Phi$'s. For a recent numerical Large $N$ study, see~\cite{Koch:2021yeb}. In this paper we will primarily use the O($N$) vector model as a concrete solvable theory. Here one has the bi-local collective field
\begin{equation}
    \Phi(t, x_1, x_2) = \frac{1}{N} \sum_{i=1}^{N} \varphi^{i}(t, x_1) \varphi^{i}(t, x_2) \, .
\end{equation}
The collective Hamiltonian is given by
\begin{equation}
    H_{\col} = \frac{2}{N} \Tr(\Pi \Phi \Pi) + \frac{N}{8} \operatorname{Tr}(\Phi^{-1}) + N V[\Phi] \, .
\end{equation}
In 3 dimensional spacetime, the O($N$) vector model has two fixed points, the UV fixed point and the Wilson-Fisher IR fixed point, at which it possesses conformal symmetry and is dual to higher spin theory in AdS${}_4$ \cite{Vasiliev:1999ba,Didenko:2014dwa,Klebanov:2002ja,Das:2003vw,Giombi:2009wh}. There is an exact map from the bi-local field $\Phi$ in O($N$) CFT to higher spin fields $\mathcal{H}$ in AdS. For example, in AdS${}_4$/CFT${}_3$ we have \cite{deMelloKoch:2014vnt}
\begin{equation}
    \mathcal{H}(\vec{p}, p^z, \theta) = \int \dd[2]{\vec p_1} \dd[2]{\vec p_2} \mathcal{K}(\vec{p}, p^z, \theta; \vec{p}_1, \vec{p}_2) \Phi(\vec p_1, \vec p_2) \, ,
\end{equation}
with the kernel $\mathcal{K}(\{\vec p\}_{\rm AdS}, \{\vec p\}_{\rm CFT})$ representing a canonical transformation of the momenta. Here $\theta$ is a coordinate in $\mathbb{S}^1$ which packages all spin variables. The above construction represents a duality in the time-like gauge, and other constructions such as the light-cone gauge~\cite{deMelloKoch:2010wdf} and the covariant case \cite{deMelloKoch:2018ivk} are also possible.

\section{Translations: the Large \texorpdfstring{$N$}{N} Soliton}
\label{Sec:Translations}

Generally the vacuum (ground state) solution is manifestly translationally invariant, namely $\Phi_0(x_1 + a, x_2 + a) = \Phi_0(x_1, x_2)$ for an arbitrary translational parameter $a$. However, in general one can also have other non-translationally symmetric solutions, such as large $N$ coherent states and solitons. Concrete examples were constructed in the nonlinear sigma model~\cite{Jevicki:1980kx} representing Large $N$ solitons. Denoting such solution as $\Phi_{s}(x_1, x_2)$, one has that the translational symmetry is broken:
\begin{equation}
    \Phi_{s}(x_1 + a, x_2 + a) \neq \Phi_{s}(x_1, x_2) \, ,
\end{equation}
since the solution does not commute with the momentum operator:
\begin{equation}
    [P, \Phi_{s}] = - \ii (\partial_{x_1} + \partial_{x_2}) \Phi_{s} \equiv - \ii \partial_{12} \Phi_{s}\, .
\end{equation}
In this case, problems arise when performing $1/N$ expansions around the backgrounds. In the naive expansion
\begin{align} \label{eq:Phis_1/N}
    \Phi(x_1, x_2) & = \Phi_{s}(x_1, x_2) + \frac{1}{\sqrt{N}} \hat{\Phi}(x_1, x_2) \, , \\
    \label{eq:Pis_1/N}
    \Pi(x_1, x_2) & = \sqrt{N} \,  \hat{\Pi}(x_1, x_2) \, .
\end{align} 
The collective Hamiltonian becomes
\begin{equation}
    H_{\col} = M_{0} + H_{\col}^{(2)}[\hat{\Pi}, \hat{\Phi}] + O(N^{-1/2}) \, .
\end{equation}
The leading term $M_0 = V_{\col}[\Phi_{s}]$ is the mass of the soliton, and it is of order $O(N)$. The $O(1)$ term $H_{\col}^{(2)}$ is quadratic:
\begin{equation}
    H_{\col}^{(2)} = \frac{1}{2} \operatorname{Tr}(\hat{\Pi} \Omega_s \hat{\Pi} + \hat{\Phi} V \hat{\Phi}) \, .
\end{equation}
In this equation $\Omega_s \equiv \Omega[\Phi_{s}]$ and $V \equiv \delta^2 V_{\col}[\Phi_{s}]/\delta \Phi^2$. In particular, we shall define the soliton state for the small fluctuations as
\begin{equation}
    \ket{s, 0} = \ket{\Phi_s(x_1, x_2)} \ee^{- \frac{1}{2} \Tr(\hat{\Phi} G^{-1} \hat{\Phi})} \, ,    
\end{equation}
with $G$ the static two-point function
\begin{equation}
    G = \sum_{n = 0}^{\infty} \frac{f_n^{*} f_n}{2 \omega_n} \, , \qquad \Omega_s V f_n = \omega_n^2 f_n \, .
\end{equation}
The notion for the state $\ket{s, x=0}$ indicates that the center of mass of the soliton is located at $x=0$. 

\paragraph{}
We would like to implement translations: 
\begin{equation}
    \ket{s, a} = \ee^{- \ii P a} \ket{s, 0} \, ,
\end{equation}
with the momentum operator that is expanded as:
\begin{align} \label{eq:momentum_operator_soliton}
    P = \Tr(\Pi \partial_{12} \Phi) & \equiv P_1 + P_2 \\
    & \equiv \sqrt{N} \operatorname{Tr}(\hat{\Pi} \partial_{12} \Phi_{s}) + \operatorname{Tr}(\hat{\Pi} \partial_{12} \hat{\Phi}) \, ,
\end{align}
where we use the shorthand notation $\partial_{12} \equiv \partial_{x_1} + \partial_{x_2}$. We note the leading term in this expansion: $P_1$ is of order $\sqrt{N}$ while $P_2$ is of order $1$. Due to this $N$ dependence of $P$, one cannot manifestly see translations in the naive large $N$ expansion scheme. For example, an infinite series re-summation is needed to evaluate
\begin{align}
    \ee^{\ii a P} \Phi(x_1 , x_2) \ee^{- \ii a P} & = \ee^{\ii a (P_1 + P_2)} (\Phi_s + N^{-1/2} \hat{\Phi}) \ee^{-\ii a (P_1 + P_2)} \nonumber \\
    & = \sum_{n=0}^{\infty} \frac{(\ii a)^n}{n!} \ad_{P_1 + P_2}^{n}(\Phi_s + N^{-1/2} \hat{\Phi}) \nonumber \\
    & = \Phi_s(x_1 + a, x_2 + a) + N^{-1/2} \hat{\Phi}(x_1 + a, x_2 + a) \, , 
\end{align}
with $\ad_{A}(B) = [A, B]$. Due to $\sqrt{N}$ of the leading operator $P_1$ terms of different orders in $1/N$ mix up. As a concrete ingredient in this transformation, let us consider $[P_1, N^{-1/2} \hat{\Phi}]$. Since $P_1$ is of order $\sqrt{N}$ and $N^{-1/2} \hat{\Phi}$ is of order $N^{-1/2}$, this gives an order $1$ term. As a result, it contributes to transformations of the background term $\Phi_s$ instead of $\hat{\Phi}$. In matrix models, the leading term $P_1$ is of order $N$, and the situation gets even worse. We will see in subsequent sections that similar issue also arises in large $N$ expansion around thermofield double states.

\paragraph{}
In addition, the presence of the zero mode frequency $\omega_0 = 0$ implies that the static two-point function $G$ is divergent and the propagator is ill-defined. The appearance of the zero mode is related to breaking of (translational) symmetry: $f_0$ is the Goldstone mode. In particular, consider the symmetry condition $[H_{\col}, P] = 0$, in the large $N$ limit of the soliton sector, this condition reduces to $[H_{\col}^{(2)}, P_1] = 0$, and yields
\begin{equation}
    V f_0 = 0 \, , \qquad f_0 = \partial_{12} \Phi_{s} \, .
\end{equation}
This results in infrared divergences, in the sense that $G$ being singular, due to the zero mode for $n=0$, hence perturbation in terms of $1/N$ is not possible.

\paragraph{}
In QFT, the canonical way to project out the zero mode and to develop a systematic perturbation expansion in terms of the coupling constant is the collective coordinate method~\cite{Gervais:1975pa,Gervais:1975yg,Christ:1975wt,Korepin:1975zu}. In that case the soliton background $\phi_{s} \sim 1/g$ with $g$ the coupling constant. In the case of the large $N$ nonlinear sigma model, the role of $g$ is played by $1/\sqrt{N}$. The collective coordinate is identified with the position of the center of mass of the soliton, denoted $\hat{x}(t)$, which is now promoted to a degree of freedom. We are led to work in the extended Hilbert space: in addition to $\hat{\Phi}$ and $\hat{\Pi}$, we would also have $\hat{x}$ and its conjugate $\hat{p}$, with $[\hat{x}, \hat{p}] = \ii$. These new variables obey constraints and gauge conditions
\begin{align}
    \label{eq:p_constraint}
    \hat{p} - P[\Pi, \Phi] \ket{s, 0} & = 0 \, , \\
    \label{eq:X_gauge_general}
    \chi_{\hat{x}}[\Pi, \Phi] \ket{s, 0} & = 0 \, .
\end{align}
The gauge condition \cref{eq:X_gauge_general} actually can be arbitrary \cite{Gervais:1975yg}. Here $\chi = \chi[\Pi, \Phi]$ denotes an arbitrary functional of $\Pi$ and $\Phi$, and $\chi_{\hat{x}} \equiv \ee^{-\ii \hat{x} P} \chi \ee^{\ii \hat{x} P}$. To project out the zero mode in the simplest way, we choose the linear gauge condition 
\begin{equation}
    \label{eq:X_gauge_simpler}
    \int f_0 \Phi(x_1 + \hat{x}, x_2 + \hat{x}) \dd{x_1} \dd{x_2} \ket{s, 0} = 0 \, .
\end{equation}
On the other hand, one can also choose the canonical gauge condition \cite{Korepin:1975zu}
\begin{equation} \label{eq:X_gauge}
    \hat{x} - \frac{\int (x_1 + x_2) \mathcal{H}_{\col} \dd{x_1} \dd{x_2}}{H_{\col}} \ket{s, 0} = 0 \, ,
\end{equation}
where $\mathcal{H}_{\col}$ is the collective Hamiltonian density. One can verify that the solutions for these equations indeed obey the canonical commutation relation. Since this gauge is solved by $\hat{x} = K / H$, where $K$ is the boost operator, using Poincar\'{e} algebra, we see that the canonical commutation relation is guaranteed 
\begin{equation}
    [\hat{x}, \hat{p}] = [\frac{K}{H}, P] = \ii \, .
\end{equation}

\paragraph{}
With this collective coordinate we are able to make a change of coordinate system, through a unitary transformation. We can perform a translation $x \rightarrow x + \hat{x}$ to the soliton frame
\begin{align}
    \Phi^{\prime}(x_1, x_2) & = \ee^{\ii \hat{x} P} \Phi(x_1, x_2) \ee^{- \ii \hat{x} P} = \Phi(x_1 + \hat{x}, x_2 + \hat{x}) \, , \\
    \Pi^{\prime}(x_1, x_2) & = \ee^{\ii \hat{x} P} \Pi(x_1, x_2) \ee^{- \ii \hat{x} P} = \Pi(x_1 + \hat{x}, x_2 + \hat{x}) \, ,
\end{align}
and similarly for all field degrees of freedom, including their $1/N$ expansions. The inverse transformations are
\begin{align} \label{eq:Phi_Phiprime_expand}
    \Phi(x_1, x_2) & = \Phi^{\prime}(x_1 - \hat{x}, x_2 - \hat{x}) = \Phi^{\prime}(x_1, x_2) - \hat{x} \partial_{12} \Phi^{\prime}(x_1, x_2) + \dots \, , \\
    \Pi(x_1, x_2) & = \Pi^{\prime}(x_1 - \hat{x}, x_2 - \hat{x}) = \Pi^{\prime}(x_1, x_2) - \hat{x} \partial_{12} \Pi^{\prime}(x_1, x_2) + \dots \, . 
\end{align}
In particular, we can apply a translation to the soliton state $\ket{s, 0}^{\prime} = \ee^{ \ii \hat{x} P} \ket{s, 0}$, such that in coordinate space, the new state becomes
\begin{equation}
    \ket{s, 0}^{\prime} = \ket{\Phi_{s}(x_1 + \hat{x}, x_2 + \hat{x})} \ee^{- \frac{1}{2} \Tr(\hat{\Phi}^{\prime} G^{\prime -1} \hat{\Phi}^{\prime})} \, ,
\end{equation}
with the zero mode projected out in $G^{\prime}$:
\begin{equation}
    G^{\prime} = \sum_{n = 1}^{\infty} \frac{f_n^{*} f_n}{2 \omega_n} \, .
\end{equation}
Then $\ket{s, 0}^{\prime}$ can be translated easily via
\begin{equation}
    \ket{s, a}^{\prime} = \ee^{ \ii a \hat{p}} \ket{s, 0}^{\prime} \, .
\end{equation}
We also have momentum eigenstates
\begin{equation}
    \ket{s, p}^{\prime} = \int \dd{a} \ee^{- \ii a p} \ket{s, a}^{\prime} \, .
\end{equation}
As illustrated above, the collective coordinate enables one to implement translations on states and fields without the expansion the momentum operator $P$. On  the extended Hilbert space, translation symmetric forms directly follow. For example, we have for the form factor (one-point function):
\begin{equation}
    \bra{s, p^{\prime}}^{\prime}  \Phi_{s}^{\prime}(x_1 - \hat{x}, x_2 - \hat{x}) \ket{s, p}^{\prime} = \int \dd{y} \ee^{\ii (p - p^{\prime}) y} \Phi_s^{\prime}(x_1 - y, x_2 - y) \, .
\end{equation}
Or in the example of the two-point function the correlation function, we have
\begin{equation}
    \langle \Phi(x_1, x_2, t) \Phi(y_1, y_2, t_0) \rangle = 
    \langle \Phi^{\prime}(x_1 - \hat{x}(t), x_2 - \hat{x}(t), t) \Phi^{\prime}(y_1 - \hat{x}(t_0), y_2 - \hat{x}(t_0), t_0) \rangle \, . 
\end{equation}

\paragraph{}
Generally the coordinate $\hat{x}(t)$ is treated as a dynamical variable. The transformation of shifting to $x_1 - \hat{x} = \rho_1$ and $x_2 - \hat{x} = \rho_2$, represents a change of frame. The constraint and the gauge condition \cref{eq:p_constraint,eq:X_gauge_simpler} then become
\begin{align}
    \hat{p} - \Tr(\Pi^{\prime} \partial_{12} \Phi^{\prime}) \ket{s, 0}^{\prime} & = 0 \, , \\
    \int f_0 \Phi^{\prime}(\rho_1, \rho_2) \dd{\rho_1} \dd{\rho_2} \ket{s, 0}^{\prime} & = 0 \, .
\end{align}
In particular, the gauge condition implies that the zero mode is now projected out, and a systematic $1/N$ expansion can be developed. Writing
\begin{equation}
    \Phi^{\prime} = \Phi^{\prime}_{s} + \frac{1}{\sqrt{N}} \hat{\Phi}^{\prime} \, , 
    \qquad
    \Pi^{\prime} = \sqrt{N} (\Pi_s^\prime + \hat{\Pi}^{\prime}) \, ,
\end{equation}
we see that the constraint becomes
\begin{equation}
     \hat{p} - \sqrt{N} \operatorname{Tr}(\Pi^{\prime}_s \partial_{12} \Phi^{\prime}) - \sqrt{N} \operatorname{Tr}(\hat{\Pi}^{\prime} \partial_{12} \Phi^{\prime}_{s}) - \operatorname{Tr}(\hat{\Pi}^{\prime} \partial_{12} \hat{\Phi}^{\prime}) \ket{s, 0}^{\prime} = 0 \, ,
\end{equation}
and with the requirement 
\begin{equation}
    \operatorname{Tr}(\hat{\Pi}^{\prime} \partial_{12} \Phi^{\prime}_{s}) \ket{s, 0}^{\prime} = 0 \, ,
\end{equation}
one solves for $\Pi'_s$ to have
\begin{equation}
    \Pi'_s=\frac{\partial_{12} \Phi^{\prime}_{s}}{\sqrt{N}} 
        \frac{
        \hat{p} - \operatorname{Tr}(\hat{\Pi}^\prime\partial_{12}\hat{\Phi}^\prime)
        }{
        \operatorname{Tr}(\partial_{12}\Phi_s^\prime \partial_{12}\Phi^\prime)
        } \, .
\end{equation}
Similarly we can expand the gauge condition, and find at leading order
\begin{equation}
    \operatorname{Tr}(\hat{\Phi}^{\prime} \partial_{12} \Phi^{\prime}_{s}) \ket{s, 0}^{\prime} = 0 \, .
\end{equation}
Thus, the zero mode is projected out from the linear fluctuation fields. 
Correspondingly, the wave functional for the small fluctuations $\hat{\Phi}^{\prime}$ around the soliton background is 
\begin{equation}
    \Psi[\hat{\Phi}^{\prime}] = \mathcal{N} \ee^{\ii p \, x} \ee^{- \frac{1}{2} \operatorname{Tr}(\hat{\Phi}^{\prime} G^{\prime \, - 1} \hat{\Phi}^{\prime})} \, ,
\end{equation}
with $G^{\prime}$ the (equal-time) two-point correlators $\langle \hat{\Phi}^{\prime} \hat{\Phi}^{\prime} \rangle$ with the zero mode excluded.

\paragraph{}
The  Hamiltonian becomes
\begin{equation}
\begin{split}
    H_{\col} =& +\frac{M_0}{2}\left(\frac{\hat{p} - \operatorname{Tr}(\hat{\Pi}^\prime\partial_{12}\hat{\Phi}^\prime)}{\operatorname{Tr}(\partial_{12}\Phi_s^\prime \partial_{12}\Phi^\prime)}\right)^2 + \frac{1}{2} \operatorname{Tr}(\hat{\Pi}^{\prime} \Omega_s \hat{\Pi}^{\prime} + \hat{\Phi}^{\prime} V \hat{\Phi}^{\prime}) \\
    &- \frac{1}{8\sqrt{N}}\operatorname{Tr}(\frac{1}{\Phi_s^{\prime}}*\hat{\Phi}^{\prime}*\frac{1}{\Phi_s^{\prime}}*\hat{\Phi}^{\prime}*\frac{1}{\Phi_s^{\prime}}*\hat{\Phi}^{\prime}*\frac{1}{\Phi_s^{\prime}}) + O(N^{-1})\, ,
\end{split}
\end{equation}
where
\begin{equation}
    M_0=\operatorname{Tr}(\partial_{12} \Phi^{\prime}_{s})^2 \, ,
\end{equation}
and can be systematically expanded in  $1/N$ as
\begin{equation}
\begin{split}
    H_{\col} = & +\frac{\hat{p}^2}{2M_0} + \frac{1}{2} \operatorname{Tr}(\hat{\Pi}^{\prime} \Omega_s \hat{\Pi}^{\prime} + \hat{\Phi}^{\prime} V \hat{\Phi}^{\prime}) \\
    & - \frac{\hat{p}}{M_0}\operatorname{Tr}(\hat{\Pi}^\prime\partial_{12}\hat{\Phi}^\prime) - \frac{\hat{p}^2}{M_0^2}\operatorname{Tr}(\partial_{12}\Phi^\prime_s\partial_{12}\hat{\Phi}^\prime) \\
    & - \frac{1}{8\sqrt{N}}\operatorname{Tr}(\frac{1}{\Phi_s^{\prime}}*\hat{\Phi}^{\prime}*\frac{1}{\Phi_s^{\prime}}*\hat{\Phi}^{\prime}*\frac{1}{\Phi_s^{\prime}}*\hat{\Phi}^{\prime}*\frac{1}{\Phi_s^{\prime}})
    + O(N^{-1}) \, ,
\end{split}
\end{equation}
providing the basis for the $1/N$ expansion.

\section{Thermofield Double State at Large \texorpdfstring{$N$}{N}}
\label{sec:TFD_state_at_large_N}


We follow the Hamiltonian formalism for the Thermofield double (TFD) \cite{Takahasi:1974zn}. The TFD state $\ket{0(\beta)}$ is introduced to completely reproduce the thermal averages of various operators:
\begin{equation}
    \langle \mathcal{O} \rangle_{\beta} \equiv \bra{0(\beta)} \mathcal{O} \ket{0(\beta)} = \frac{1}{Z(\beta)} \operatorname{Tr}(\ee^{-\beta H} \mathcal{O}) \, .
\end{equation}
This is achieved through purification by doubling the Hilbert space. Let $\ket{\tilde{n}}$ denote the energy eigenstates in the doubled Hilbert space with $\tilde{n} = n$,
\begin{equation}
    \ket{0(\beta)} = \frac{1}{\sqrt{Z(\beta)}} \sum_{n} \ee^{- \beta E_{n} / 2} \ket{n} \ket{\tilde{n}} \, ,
\end{equation}
In the context of AdS/CFT correspondence, such states involves two identically copies of large $N$ CFTs at two causally disconnected boundaries, and are dual to two-sided eternal black holes in the bulk \cite{Maldacena:2001kr}. The Hamiltonian that governs the dynamics is
\begin{equation}
    \hat{H} = H - \tilde{H} \, .
\end{equation}
Obviously it annihilates the TFD state for all $\beta$, i.e. $\hat{H} \ket{0(\beta)} = 0$. The thermofield Hamiltonian $\hat{H}$ describes the real time portion of the Schwinger-Keldysh contour, and the TFD state can be referred to as the thermal vacuum state. 

\paragraph{}
Evolution along the imaginary portion of the Schwinger-Keldysh contour is governed by the Hamiltonian
\begin{equation}
    H_{+} = H + \tilde{H} \, ,
\end{equation} 
and we have
\begin{equation} \label{eq:Hplus_transformation}
    \ket{0(\beta)} = \frac{1}{\sqrt{Z(\beta)}} \ee^{- \beta H_{+} / 4} \ket{I} \, , 
\end{equation}
where 
\begin{equation}
    \ket{I} \equiv \sqrt{Z(0)} \ket{0(0)} = \sum_{n} | n \rangle | \tilde{n} \rangle
\end{equation}
denotes the maximally entangled state. This relates $\ket{0(\beta)}$ with the infinite temperature state $\ket{I}$ through a non-unitary transformation. 
Most importantly one has that the two Hamiltonians commute (representing a symmetry):
\begin{equation}
    [\hat{H}, H_{+}] = 0 \, .
\end{equation}

\paragraph{} 
A further dynamical symmetry has been argued at the semiclassical level (in the sense of Large $N$) in \cite{Jevicki:2021ddf}. In the free O($N$) model case we indeed have
\begin{equation} \label{eq:G2}
    \hat{G} = \int \theta(\vec k) \hat{\mathcal{G}}(\vec k) \dd[d]{\vec k} \, , 
    \qquad 
    \hat{\mathcal{G}}(\Vec{k}) =  \ii  \Big(a^{\dagger i}(\vec{k}) \widetilde{a}^{\dagger i}(\vec{k}) - a^{i}(\vec{k}) \widetilde{a}^{i}(\vec{k})\Big) \, , 
    \qquad
    \tanh \theta(\vec k) = \ee^{- \beta \omega(\vec k)/2} \, ,
\end{equation}
with
\begin{equation} \label{eq:Bogoliubov_operators}
    \mathcal{O}_{\theta} := \ee^{- \ii \hat{G}} \mathcal{O} \ee^{\ii \hat{G}}
\end{equation}
representing a Bogoliubov transformation. 
In \cite{Jevicki:2021ddf} a construction of $\hat{G}$ to first order in the coupling is shown. Generally this appears to be an on shell symmetry, which however will play a role at the level of fluctuations (in $1/N$).

\paragraph{}
We continue with the  O($N$) model 
\begin{equation}
    H[\pi, \varphi] = \int \left[ \frac{1}{2} \pi^{i} \pi^{i} + \frac{1}{2} \nabla \varphi^i \nabla \varphi^i + \frac{m^2}{2}  \varphi^i \varphi^i + \frac{c}{4 N} (\varphi^{i} \varphi^{i})^2 \right] \dd^{d} \vec{x} \, ,
\end{equation}
which at UV and IR critical points represents the CFT. At finite temperature, the model has a phase transition \cite{Shenker:2011zf} with free energy $F(T) \sim N T^2$ for $T>T_c$ and $F(T) \sim T^4$ for $T<T_c$. For the TFD scheme, we also have $\tilde{H}$ with $\tilde{\varphi}$ and $\tilde{\pi}$, with the theory generally has an $\operatorname{O}(N) \times \operatorname{O}(N)$ symmetry. The following structure regarding gauging of the O($N$) symmetry was seen in \cite{Jevicki:2015sla}.

\paragraph{}
For the lower temperature (the AdS-phase) one imposes the singlet constraint on the original and the doubled Hilbert spaces, namely
\begin{equation}
    J^{i j}\ket{\Phi} = 0 \, , \qquad \Tilde{J}^{i j}\ket{\Phi} = 0 \, ,
\end{equation}
where $J^{ij}$ and $\Tilde{J}^{ij}$ are $O(N)$ generators of $\varphi$ and $\Tilde{\varphi}$ . This implies that we have two invariant bi-local (only in space) fields in the spectrum
\begin{equation}
\label{eq:bi-local}
    \Phi^{11}(t; \vec{x}_1, \vec{x}_2) = \frac{1}{N}\varphi^{i}(t, \vec{x}_1) \varphi^{i}(t, \vec{x}_2) \, , \qquad 
    \Phi^{22}(t; \vec{x}_1, \vec{x}_2) = \frac{1}{N}\tilde{\varphi}^{i}(t, \vec{x}_1) \tilde{\varphi}^{i}(t, \vec{x}_2) \, ,
\end{equation}
representing a direct product Hilbert spaces of $\rm{CFT} \times \tilde{\rm{CFT}}$.

\paragraph{}
For the high temperature phase, it was proposed in \cite{Jevicki:2015sla} that one needs to relax the above constraints and has diagonal gauging gauging of O($N$)   
\begin{equation}
    (J^{i j} + \Tilde{J}^{i j}) \ket{\Phi} = 0 \, .
\end{equation}
Now, in addition to \cref{eq:bi-local}, we have two more bi-local fields (cross modes) in the Hilbert space
\begin{equation}
    \Phi^{12}(t; \vec{x}_1, \vec{x}_2) = \frac{1}{N}\varphi^{i}(t, \vec{x}_1) \tilde{\varphi}^{i}(t, \vec{x}_2)\, , \qquad 
    \Phi^{21}(t; \vec{x}_1, \vec{x}_2) = \frac{1}{N}\tilde{\varphi}^{i}(t, \vec{x}_1) \varphi^{i}(t, \vec{x}_2) \, .
\end{equation}
The diagonal gauging  \cite{Jevicki:2015sla} was seen to allow an order $N$ free energy at the leading classical (in the sense of $1/N$) level. At the level of fluctuations \cite{Jevicki:2015sla} the cross modes were seen to be responsible for the presence of evanescent modes and for generating a complete spectrum in the bulk.
The presence of cross modes implies that one can not have a direct product of two CFTs, since they interact here through these mixed modes.
We note that at the gravity level various issues have been discussed for the  $\operatorname{CFT}_L \times \operatorname{CFT}_R$ two-sided black hole duality scheme in particular in \cite{Mathur:2012dxa,Marolf:2012xe,Jensen:2014lua}. Also Re: gauging in CFT one has a parallel in recent gravitational studies (for two-sided wormhole space-times)  with diagonal implementation of constraint symmetries \cite{Jafferis:2019wkd,Penington:2023dql,Engelhardt:2022qts}.

\paragraph{}
Consequently, we use the bi-local collective fields\footnote{
    A bit explanation of the notations: Here $\Phi^{ab}(\vec{x}, \vec{y}) \equiv \varphi^{i}_{a}(\vec{x}) \varphi^{i}_{b}(\vec{y})/N$ with $\varphi^{i}_{1} = \varphi$ and $\varphi^{i}_{2} = \tilde{\varphi}^{i}$. In the context of AdS/CFT, one also denotes $ \varphi_{R}^{i} \equiv \varphi^{i}$ and $\varphi_{L}^{i} \equiv \tilde{\varphi}^{i}$.
}
\begin{equation}
    \Phi(\vec{x}, \vec{y}) \equiv \begin{pmatrix}
        \Phi^{11} & \Phi^{12} \\ 
        \Phi^{21} & \Phi^{22} 
    \end{pmatrix} 
    (\vec{x}, \vec{y})
    := \frac{1}{N} \begin{pmatrix}
        \varphi^i \varphi^i & \varphi^{i} \widetilde{\varphi}^{i} \\
        \widetilde{\varphi}^i \varphi^i & \widetilde{\varphi}^i \widetilde{\varphi}^i
    \end{pmatrix} (\vec{x}, \vec{y}) 
    \, ,
\end{equation}
and their canonical conjugates $\Pi = - \ii \delta / \delta \Phi$, to represent the thermofield Hamiltonian $\hat{H}$ as $\widehat{H}_{\col}[\Pi, \Phi]$, namely $\hat{H}[\pi, \tilde{\pi}, \varphi, \tilde{\varphi}] = \hat{H}_{\col}[\Pi, \Phi]$ as \cite{Jevicki:2015sla,Jevicki:2021ddf}
\begin{align}
    \hat{H}_{\col} = & \frac{2}{N} \Tr[\Pi \star (\sigma_3 \Phi) \star \Pi] + \frac{N}{8} \Tr[\sigma_3 \Phi^{-1}] + \frac{N}{2} \Tr[(-\nabla^2 + m^2) \star(\sigma_3 \Phi)] \nonumber \\
    & + \frac{N c}{4}\int \left\{ [\Phi^{11}(\vec{x}, \vec{x})]^2 - [\Phi^{22}(\vec{x}, \vec{x})]^2 \right\} \dd[d]{\vec{x}} \, ,
\end{align}
with $\sigma_3 = \operatorname{diag}(1, -1)$ the Pauli third matrix, and $\Tr(A \star B) \equiv \int A(\vec{x}, \vec{y}) B(\vec{y}, \vec{x}) \dd[d]{\vec{x}} \dd[d]{\vec{y}}$. 
To obtain the large $N$ thermal background, we vary $\hat{H}_{\col}$ with respect to $\Phi$:
\begin{equation}
    \fdv{\hat{H}_{\col}}{\Phi} = 0 \, ,
\end{equation}
which gives the equation determining the large $N$ thermal background. Several important features of this (classical) Large $N$ equation were identified in \cite{Jevicki:2015sla,Jevicki:2021ddf} foremost being the appearance of a symmetry (mentioned above). Namely the equation allows for 
\begin{equation}
    \Phi_{f} (\vec{x}, \vec{y}) = \int \frac{\dd[d]{\vec{k}}}{(2\pi)^d} \frac{\ee^{\ii \vec{k} \cdot (\vec{x} - \vec{y})}}{2 \omega_{f}(\vec{k})} 
    \begin{pmatrix}
        \cosh f(\vec{k}) & \sinh f(\vec{k}) \\
        \sinh f(\vec{k}) & \cosh f(\vec{k})
    \end{pmatrix} \, ,
\end{equation}
a one-parameter family of solutions, with $f$ representing the parameter. The classical solution corresponds to the thermal two-point function at equal time, namely $\Phi^{ab}_{\theta}(\vec{x}, \vec{y}) = \langle \varphi_{a}^{i}(t, \vec{x}) \varphi_{b}^{i}(t, \vec{y}) \rangle_{\beta} / N$ and  one then identifies $f(\vec{k}) = 2 \theta(\vec{k})$. The solution is manifestly translational invariant, as in the zero temperature case. The dispersion relation $\omega_f(\vec k)$ obeys a thermal gap equation \cite{Jevicki:2021ddf}. 
It was understood in the previous work \cite{Jevicki:2021ddf} that this degeneracy can be attributed to $\hat{G}$-symmetry, and the large $N$ thermal background is related to the zero temperature background $\Phi_0$ via (note that the signs are opposite to those in \cref{eq:Bogoliubov_operators})
\begin{equation}
    \Phi_{f} \equiv \left. \ee^{\ii \hat{G}_{f}} \Phi \ee^{-\ii \hat{G}_{f}} \right|_{\Phi = \Phi_{0}} \, .
\end{equation}
$\hat{G}$ in general appears to be an on shell symmetry, in particular in the general interacting case. Consequently it manifests itself in the semi-classical approximation (in the sense of 1/$N$) as above. It will also manifest itself at the level of fluctuations .

\paragraph{}
We emphasize that this property is not limited to the O($N$) vector models, but also applies to more complicated theories, such as matrix quantum mechanics at finite temperature (to be presented in the future work).

\paragraph{}
For simplicity in the following we will consider the free massless theory case. One performs a shift around the thermal background
\begin{equation}
    \Pi \rightarrow \sqrt{N} \pi \, , 
    \qquad
    \Phi \rightarrow \Phi_{\theta} + \frac{1}{\sqrt{N}} \eta \, .
\end{equation}
We will write $\eta$ and its conjugate $\pi$ as vectors whose components are bi-local fields in momentum space:
\begin{equation}
    \boldsymbol \pi = \begin{pmatrix}
        \pi^{11} \\
        \pi^{12} \\
        \pi^{21} \\
        \pi^{22}
    \end{pmatrix} \, , 
    \qquad
    \boldsymbol \eta = \begin{pmatrix}
        \eta^{11} \\
        \eta^{12} \\
        \eta^{21} \\
        \eta^{22}
    \end{pmatrix} \, .
\end{equation}
The collective Hamiltonian has a systematic $1/N$ expansion:
\begin{equation}
    \widehat{H}_{\col}[\boldsymbol \pi, \boldsymbol \eta] = \sum_{n=0}^{\infty} N^{1 - \frac{n}{2}} \widehat{H}_{\col}^{(n)}[\boldsymbol \pi, \boldsymbol \eta] 
    = \hat{H}_{\col}^{(2)} + \frac{1}{\sqrt{N}} \hat{H}_{\col}^{(3)} + \dots \, .
\end{equation}
This implies that
\begin{equation}
    \widehat{H}_{\col}^{(0)} = \widehat{H}_{\col}^{(1)} = 0 \, .
\end{equation}
The first equality states that in the strict large $N$ limit, $\hat{H}$ must annihilate the TFD state. The second one states that thermal backgrounds correspond to the saddle point solutions of $\hat{H}$. At order 1 (i.e. $n=2$), we have a quadratic form (The trace is taken in the bi-local momentum space.)
\begin{equation} \label{eq:Hcol_2}
    \widehat{H}_{\col}^{(2)} = \frac{1}{2} \operatorname{Tr}[\boldsymbol{\pi}^{\operatorname{T}} K \boldsymbol{\pi} + \boldsymbol{\eta}^{\operatorname{T}} V \boldsymbol{\eta}] \, ,
\end{equation}
The kinetic matrix $K$ is given by
\begin{equation} \label{eq:K}
    K(\vec{k}_1, \vec{k}_2) = 
    \begin{pmatrix}
        c_1 + c_2 & s_2 & s_1 & 0 \\
        s_2 & - c_1 + c_2 & 0 & - s_1 \\
        s_1 & 0 & c_1 - c_2 & - s_2 \\
        0 & - s_1 & - s_2 & - c_1 - c_2
    \end{pmatrix} \, ,
\end{equation}
where
\begin{equation} \label{eq:c_s}
    c_i \equiv \frac{\cosh 2 \theta(\vec{k}_i)}{\omega(\vec{k}_i)} \, , \qquad
    s_i \equiv \frac{\sinh 2 \theta(\vec{k}_i)}{\omega(\vec{k}_i)} \, ,
\end{equation}
The potential matrix is $V$ is given by
\begin{equation} \label{eq:V}
    V(\vec{k}_1, \vec{k}_2) =
    \omega^{2}(\vec{k}_1) \omega^{2}(\vec{k}_2)
    \begin{pmatrix}
        c_1 + c_2 & - s_2 & - s_1 & 0 \\
        - s_2 & - c_1 + c_2 & 0 & s_1 \\
        - s_1 & 0 & c_1 - c_2 & s_2 \\
        0 & s_1 & s_2 & - c_1 -c_2
    \end{pmatrix} \, .
\end{equation}

\paragraph{}
We now turn to the TFD wave functional $\Psi_{\beta}[\boldsymbol{\eta}]$ which  is an  eigenstate of $\hat{H}_{\col}^{(2)}$: 
\begin{equation} \label{eq:Hhat_Psi_is_0}
    \hat{H}_{\col}^{(2)} \Psi_{\beta}[\boldsymbol{\eta}] = 0 \, .
\end{equation}
Based on our previous results \cite{Jevicki:2021ddf}, we have :
\begin{equation} \label{eq:TFD_wave_function_eta}
    \Psi_{\beta}[\boldsymbol{\eta}] = \mathcal{N} \exp[
        - \frac{1}{2} \int \boldsymbol{\eta}^{\operatorname{T}}(\vec{k}_1, \vec{k}_2) \, G^{-1}(\vec{k}_1, \vec{k}_2) \, \boldsymbol{\eta}(\vec{k}_1, \vec{k}_2) \frac{\dd[d]{\vec{k}_1}}{(2\pi)^d} \frac{\dd[d]{\vec{k}_2}}{(2\pi)^d}
    ] \, ,
\end{equation}
where one has the equal-time two-point functions of $\eta$'s at finite temperature 
\begin{equation} \label{eq:G_correlator_relation}
    G^{ab, cd}(\vec{k}_1, \vec{k}_2) = \int \frac{\dd[d]{\vec{k}_3}}{(2\pi)^{d}} \frac
    {\dd[d] \vec{k}_4}{(2\pi)^{d}} \langle \eta^{ab}(\vec{k}_1, \vec{k_2}) \eta^{cd}(\vec{k}_3, \vec{k}_4) \rangle_{\beta} \, .
\end{equation}
Explicitly (at $c=0$),
\begin{equation}
    G^{-1}(\vec{k}_1, \vec{k}_2) = \omega^2(\vec{k}_1) \omega^2(\vec{k}_2)
    \begin{pmatrix}
        c_1 c_2 & - c_1 s_2 & - s_1 c_2 & s_1 s_2 \\
        - c_1 s_2 & c_1 c_2 & s_1 s_2 & - s_1 c_2 \\
        - s_1 c_2 & s_1 s_2 & c_1 c_2 & - c_1 s_2 \\
        s_1 s_2 & - s_1 c_2 & - c_1 s_2 & c_1 c_2 
    \end{pmatrix} \, .
\end{equation}

\paragraph{}
However, it should be stressed that the solution for $\Psi_{\beta}$ is not unique. As discussed in \cite{Jevicki:2021ddf}, the non-uniqueness is related to the singular structure of $\hat{H}_{\col}^{(2)}$, i.e. its zero modes. This singular structure, and the zero modes appear to be related to the symmetry pointed out above.

\subsection{Normal modes and bulk fields}

In general there exists a basis (normal modes) such that the quadratic Hamiltonian is diagonalized. In particular continuing with the 
$c=0$ case both $K$ and $V$ can be simultaneously diagonalized through a linear transformation ($\theta_{a} \equiv \theta(\vec{k}_{a})$)
\begin{equation}  \label{eq:eta_theta}
    \boldsymbol{\eta}_{\theta}(\vec{k}_1, \vec{k}_2) = M[-\theta_1, - \theta_2] \boldsymbol{\eta}(\vec{k}_1, \vec{k}_2) \, .
\end{equation}
Let
\begin{equation}
    \mathfrak{c}_i \equiv \cosh \theta(\vec{k}_i) \, , 
    \qquad
    \mathfrak{s}_i \equiv \sinh \theta(\vec{k}_i)\, ,
\end{equation}
$M[\theta(\vec{k}_1), \theta(\vec{k}_2)]$ is a matrix given by
\begin{equation}
    M[\theta_1, \theta_2] = 
    \begin{pmatrix}
        \mathfrak{c}_1 \mathfrak{c}_2 & \mathfrak{c}_1 \mathfrak{s}_2 &  \mathfrak{c}_2 \mathfrak{s}_1 &  \mathfrak{s}_1 \mathfrak{s}_2 \\
        \mathfrak{c}_1 \mathfrak{s}_2 &  \mathfrak{c}_1 \mathfrak{c}_2 &  \mathfrak{s}_1 \mathfrak{s}_2 &  \mathfrak{c}_2 \mathfrak{s}_1 \\
        \mathfrak{c}_2 \mathfrak{s}_1 &  \mathfrak{s}_1 \mathfrak{s}_2 &  \mathfrak{c}_1 \mathfrak{c}_2 &  \mathfrak{c}_1 \mathfrak{s}_2 \\
        \mathfrak{s}_1 \mathfrak{s}_2 &  \mathfrak{c}_2 \mathfrak{s}_1 &  \mathfrak{c}_1 \mathfrak{s}_2 &  \mathfrak{c}_1 \mathfrak{c}_2
    \end{pmatrix} \, .
\end{equation}
The $M$ matrix preserves the canonical commutation relations and generates a two-(functional)-parameter group. Its inverse thus is given by $M^{-1}[\theta_1, \theta_2] = M[- \theta_1, -\theta_2]$. It is also a symmetric matrix so that $M = M^{\operatorname{T}}$. Thus, the canonical conjugate $\boldsymbol{\pi}$ transforms as
\begin{equation} \label{eq:M_transformation_pi}
    \boldsymbol{\pi}_{\theta}(\vec{k}_1, \vec{k}_2) = M[\theta_1, \theta_2] \boldsymbol{\pi}(\vec{k}_1, \vec{k}_2) \, .
\end{equation}
For a comprehensive summary we refer the reader to \cref{appendix:Bogoliubov_transformations}.

\paragraph{}
The importance of $M$ is that it diagonalizes the matrices $K$ \eqref{eq:K} and $V$ \eqref{eq:V} \emph{simultaneously}. For $K$ we have
\begin{equation}
    M[-\theta_1, -\theta_2] K(\vec{k}_1, \vec{k}_2) M[-\theta_1, -\theta_2]^{\operatorname{T}} = K_0(\vec{k}_1, \vec{k}_2) \, , 
\end{equation}
with
\begin{equation} \label{eq:K0}
    K_0(\vec{k}_1, \vec{k}_2) = \frac{1}{\omega_1 \omega_2}
    \begin{pmatrix}
        \omega_1 + \omega_2 & & & \\
        & \omega_1 - \omega_2 & & \\
        & & - \omega_1 + \omega_2 & \\
        & & & - \omega_1 - \omega_2
    \end{pmatrix} \, .
\end{equation}
For $V$ we have
\begin{equation}
    M[\theta_1, \theta_2]^{\operatorname{T}} V(\vec{k}_1, \vec{k}_2) M[\theta_1, \theta_2] 
    = V_0(\vec{k}_1, \vec{k}_2) \, ,
\end{equation}
with
\begin{equation} \label{eq:V0}
    V_0(\vec{k}_1, \vec{k}_2) = \omega_1 \omega_2 
    \begin{pmatrix}
        \omega_1 + \omega_2 & & & \\
        & \omega_1 - \omega_2 & & \\
        & & - \omega_1 + \omega_2 & \\
        & & & - \omega_1 - \omega_2
    \end{pmatrix} \, .
\end{equation}

\paragraph{}
With this simultaneous diagonalization, we can represent the free thermofield Hamiltonian \cref{eq:Hcol_2} in terms of $\boldsymbol{\pi}_{\theta}$ and $\boldsymbol{\eta}_{\theta}$. Both $K$ and $V$ have zero modes as we will see in the following section. To illustrate the singular behaviors, it is necessary to separate 
\begin{equation} \label{eq:Hhat_separation}
    \hat{H}_{\col}^{(2)} = \hat{H}_{\col, \rm{ns}}^{(2)} + \hat{H}_{\col, \rm{s}}^{(2)} \, .    
\end{equation}
The non-singular part is
\begin{equation} \label{eq:H_col_ns}
    \hat{H}_{\col, \rm{ns}}^{(2)} = \frac{1}{2} \int_{| \vec{k}_1 | \neq | \vec{k}_2 |} \Big(
        \boldsymbol{\pi}_{\theta}^{\operatorname{T}} K_0 \boldsymbol{\pi}_{\theta} + \boldsymbol{\eta}_{\theta}^{\operatorname{T}} V_0 \boldsymbol{\eta}_{\theta}
    \Big) (\vec{k}_1, \vec{k}_2) \frac{\dd[d]{\vec{k}_1}}{(2\pi)^{d}} \frac{\dd[d]{\vec{k}_2}}{(2\pi)^{d}} \, ,
\end{equation}
and the singular part is
\begin{align} \label{eq:H_col_s}
    \hat{H}_{\col, \rm{s}}^{(2)} = \frac{1}{2} \int_{| \vec{k}_1 | = | \vec{k}_2 |} (\omega_1 + \omega_2) & \Big( 
        \frac{1}{\omega_1 \omega_2} [\pi_{\theta}^{11}(\vec{k}_1, \vec{k}_2)]^2 + 
        \omega_1 \omega_2 [\eta_{\theta}^{11}(\vec{k}_1, \vec{k}_2)]^2 \nonumber \\   
        & - \frac{1}{\omega_1 \omega_2} [\pi_{\theta}^{22}(\vec{k}_1, \vec{k}_2)]^2 - 
        \omega_1 \omega_2 [\eta_{\theta}^{22}(\vec{k}_1, \vec{k}_2)]^2
    \Big)
    \frac{\dd[d]{\vec{k}_1}}{(2\pi)^{d}} \frac{\dd[d]{\vec{k}_2}}{(2\pi)^{d}} \, .
\end{align}
Both terms are diagonal. We see that $\eta_{\theta}^{12}(\vec{k_1}, \vec{k}_2)$, $\eta_{\theta}^{21}(\vec{k}_1, \vec{k}_2)$ and their canonical conjugates with $| \vec{k}_1 | = | \vec{k}_2 |$ are completely absent from $\hat{H}_{\col}^{(2)}$. 

\paragraph{}
One also has the corresponding decomposition of the TFD wave functional \eqref{eq:TFD_wave_function_eta} :
\begin{equation}
    \Psi_{\beta}[\boldsymbol{\eta}] = \Psi_{\beta}^{\rm{ns}}[\boldsymbol{\eta}] \, \Psi_{\beta}^{\rm{s}}[\boldsymbol{\eta}] \, \Psi_{\beta}^{\rm{c}}[\boldsymbol{\eta}] \, ,
\end{equation}
where $\Psi_{\beta}^{\rm{ns}}$ and $\Psi_{\beta}^{\rm{s}}$ are Gaussian forms associated with $\hat{H}_{\col, \rm{ns}}^{(2)}$ and $\hat{H}_{\col, \rm{s}}^{(2)}$, respectively. $\Psi_{\beta}^{\rm{c}}$ is the wave functional of the missing modes $\eta^{12}_{\theta} (\vec k_1, \vec k_2)$ with $|\vec k_1| = |\vec k_2|$. They are the zero modes of $\hat{H}_{\col}^{(2)}$ and will be seen to associated with symmetry operators. For notational simplicity we will denote $\eta_0 \equiv \eta_{\theta}^{12}$ and $\pi_0 \equiv \pi_{\theta}^{12}$. Since $\hat{H}_{\col, \rm{c}}^{(2)} = 0$  we have that $\Psi_{\beta}^{\rm{c}}$ is not determined by \cref{eq:Hhat_Psi_is_0}. A full understanding of it will be seen in the framework of \Cref{sec:colcoord}. Details of the decomposition are given for completeness in the appendix \cref{appendix:TFD_state_decomposition}.

\paragraph{}
The normal modes of the linearized thermofield Hamiltonian $\hat{H}_{\col}$ are closely related to bulk free fields. In the following we discuss in detail the construction of these bulk fields $\mathcal{A}_\theta$ and $\tilde{\mathcal{A}}_\theta$. We emphasize that these bulk fields are found to be in one-to-one relationship with the collective fields. As such they should not be confused with the often used generalized free fields which are boundary CFT operators. The collective field degrees of freedom provide both the bulk wave functions and the creation-annihilation operators in the bulk. To be more explicit, $\mathcal{A}_\theta$ consists of bi-local annihilation operators $\alpha_\theta$ and $\gamma_\theta$ (to be seen in below), while $\tilde{\mathcal{A}}_\theta$ consists of bi-local annihilation operators $\tilde{\alpha}_\theta$ and $\gamma_\theta$. We would like to emphasize that these bulk fields $\mathcal{A}_\theta$ and $\tilde{\mathcal{A}}_\theta$ can not be simply interpreted as boundary operators purely from either the left or the right CFT, because the bi-local operator $\gamma_{\theta}$ (cross modes) involves both of the left and the right CFT boundary operators. To avoid notation clutters, we will restrict our attention to 3d CFT, but higher dimensional CFTs follow similarly. 

\paragraph{}
Let's introduce mode expansions for bi-local fields $\boldsymbol{\eta}_\theta$ and $\boldsymbol{\pi}_\theta$ as follows
\begin{align}
    \boldsymbol{\eta}_\theta(\vec{p}_1, \vec{p}_2) = \frac{1}{\sqrt{2|\vec{p}_1||\vec{p}_2|}}
    \begin{pmatrix}
        \alpha_\theta(\vec{p}_1,-\vec{p}_2) \\
        \gamma_\theta(\vec{p}_1,\vec{p}_2) \\
        \tilde{\gamma}_\theta(-\vec{p}_1,-\vec{p}_2) \\
        \tilde{\alpha}_\theta(-\vec{p}_1,\vec{p}_2)
    \end{pmatrix} + \operatorname{h.c.} \, , \\
    \boldsymbol{\pi}_\theta(\vec{p}_1, \vec{p}_2) = -\ii\sqrt{\frac{|\vec{p}_1||\vec{p}_2|}{2}}
    \begin{pmatrix}
        \alpha_\theta(\vec{p}_1,-\vec{p}_2) \\
        \gamma_\theta(\vec{p}_1,\vec{p}_2) \\
        \tilde{\gamma}_\theta(-\vec{p}_1,-\vec{p}_2) \\
        \tilde{\alpha}_\theta(-\vec{p}_1,\vec{p}_2)
    \end{pmatrix} + \operatorname{h.c.} \, .
\end{align}
To avoid not over-counting modes, we shouldn't integrate over the entire $\mathbb{R}^2$ which is spanned by $\vec{p}_1$ and $\vec{p}_2$. Instead, for a given $\vec{p}_2$, the integration region for $\vec{p}_1$ is $\mathbb{D}^2$ with radius $|\vec{p}_2|$ and boundary antipodally identified, which is homeomorphic to $\mathbb{RP}^2$. More explicitly,
\begin{equation}
    \left\{ (\vec{p}_1, \vec{p}_2) \Big| |\vec{p}_1| < |\vec{p}_2| \cup \left(|\vec{p}_1| = |\vec{p}_2| \cap \arcsin{\frac{\vec{p}_1 \times \vec{p}_2}{|\vec{p}_1||\vec{p}_2|}} \in [0,\pi) \right) \right\} \, .
\end{equation}
For simplicity, we abuse our notations and denote this integration region by $\mathbb{RP}^2$.

\paragraph{}
Then the quadratic Hamiltonian can be rewritten as
\begin{equation}
\begin{split}
    \hat{H}_{\col}^{(2)} =& \int_{\mathbb{RP}^2} \frac{d\vec{p}_1}{(2\pi)^2} \frac{d\vec{p}_2}{(2\pi)^2}~ (|\vec{p}_1|+|\vec{p}_2|)\left[\alpha_\theta^\dagger(\vec{p}_1,-\vec{p}_2)\alpha_\theta(\vec{p}_1,-\vec{p}_2) - \tilde{\alpha}_\theta^\dagger(-\vec{p}_1,\vec{p}_2)\tilde{\alpha}_\theta(-\vec{p}_1,\vec{p}_2)\right]\\
    & + \int_{\mathbb{RP}^2} \frac{d\vec{p}_1}{(2\pi)^2} \frac{d\vec{p}_2}{(2\pi)^2}~  (|\vec{p}_1|-|\vec{p}_2|)\left[\gamma_\theta^\dagger(\vec{p}_1,\vec{p}_2)\gamma_\theta(\vec{p}_1,\vec{p}_2) - \tilde{\gamma}_\theta^\dagger(-\vec{p}_1,-\vec{p}_2)\tilde{\gamma}_\theta(-\vec{p}_1,-\vec{p}_2)\right] \, .
\end{split}
\end{equation}
With the mass-shell condition
\begin{equation}
   k_z^2 = E^2 - |\vec{k}|^2 \, ,
\end{equation}
and the Jacobian
\begin{equation}
    J(\vec{p}_1,\vec{p}_2) = \frac{\sqrt{2|\vec{p}_1||\vec{p}_2|-2\vec{p}_1\cdot \vec{p}_2}}{|\vec{p}_1||\vec{p}_2|} \, ,
\end{equation}
the bulk fields are defined by
\begin{align}
\label{eq:gff}
    A_\theta(E,\vec{k},\phi)=&\int_{\mathbb{RP}^2} \dd \vec{p}_1 \dd \vec{p}_2~ J^{1/2}(\vec{p}_1,\vec{p}_2) \delta(|\vec{p}_1|+|\vec{p}_2|-E) \delta(\vec{p}_1+\vec{p}_2-\vec{k})\nonumber\\
    &\hspace{4cm}\times \delta\left(\arctan\frac{2\vec{p}_1\times \vec{p}_2}{(|\vec{p}_1|-|\vec{p}_2|)k_z} - \phi\right) \alpha_\theta(\vec{p}_1,-\vec{p}_2) \, ,\\
    C_\theta(E,\vec{k},\phi)=&\int_{\mathbb{RP}^2}  \dd \vec{p}_1 \dd \vec{p}_2~ J^{1/2}(\vec{p}_1,\vec{p}_2) \delta(-|\vec{p}_1|+|\vec{p}_2|-E)\delta(-\vec{p}_1+\vec{p}_2-\vec{k})\nonumber\\
    &\hspace{4cm}\times\delta\left(\arctanh\frac{2\vec{p}_1\times \vec{p}_2}{(|\vec{p}_1|+|\vec{p}_2|)k_z} - \phi\right) \gamma_\theta(\vec{p}_1,\vec{p}_2) \, ,\\
    \tilde{A}_\theta(E,\vec{k},\phi)=&\int_{\mathbb{RP}^2} \dd \vec{p}_1 \dd \vec{p}_2~ J^{1/2}(\vec{p}_1,\vec{p}_2) \delta(|\vec{p}_1|+|\vec{p}_2|-E) \delta(\vec{p}_1+\vec{p}_2-\vec{k}) \nonumber\\
    &\hspace{4cm}\times\delta\left(\arctan\frac{2\vec{p}_1\times \vec{p}_2}{(|\vec{p}_1|-|\vec{p}_2|)k_z} - \phi\right)  \tilde{\alpha}_\theta(-\vec{p}_1,\vec{p}_2) \, ,\\
    \tilde{C}_\theta(E,\vec{k},\phi)=&\int_{\mathbb{RP}^2} \dd \vec{p}_1 \dd \vec{p}_2~ J^{1/2}(\vec{p}_1,\vec{p}_2) \delta(-|\vec{p}_1|+|\vec{p}_2|-E)\delta(-\vec{p}_1+\vec{p}_2-\vec{k}) \nonumber\\
    &\hspace{4cm}\times\delta\left(\arctanh\frac{2\vec{p}_1\times \vec{p}_2}{(|\vec{p}_1|+|\vec{p}_2|)k_z} - \phi\right) \tilde{\gamma}_\theta(-\vec{p}_1,-\vec{p}_2) \, .
\end{align}
Notice that the integration region guarantees that the energy $E$ is bounded from below by $0$. We would like to emphasise that $A_\theta$ and $\Tilde{A}_\theta$ are time-like or light-like operators, while $C_\theta$ and $\Tilde{C}_\theta$ are space-like operators. It is the latter that corresponds to evanescent modes and consists of soft modes with zero energy and non-zero momentum. Collectively we denote both $A_\theta$ and $C_\theta$ as
\begin{equation}
    \mathcal{A}_\theta (E,\vec{k},\phi) = 
    \left\{\begin{array}{rcl}
        A_\theta(E,\vec{k},\phi) & \mbox{for} & E^2 \geq |\vec{k}|^2 \\
        C_\theta(E,\vec{k},\phi) & \mbox{for} & 0 \leq E^2<|\vec{k}|^2
    \end{array}\right. \,
\end{equation}
and similarly for $\Tilde{\mathcal{A}}_\theta$. Here we have the extensive field with energy $E$ ranging in $[0,\infty)$. Notice that the zero modes $\eta_0$ and $\pi_0$ in the boundary exactly correspond to the soft modes in the bulk.
The bulk free fields satisfy the regular commutation relations
\begin{equation}
    [\mathcal{A}_\theta(E,\vec{k},\phi),\mathcal{A}_\theta^\dagger(E',\vec{k}',\phi')]=[\tilde{\mathcal{A}}_\theta(E,\vec{k},\phi),\tilde{\mathcal{A}}_\theta^\dagger(E',\vec{k}',\phi')]=\delta(E-E')\delta^{(2)}(\vec{k}-\vec{k}')\delta(\phi-\phi') \, ,
\end{equation}
which implies that the spectrum is complete.
In terms of the bulk free fields, the quadratic Hamiltonian can be recast into the form
\begin{align}
\label{eq:H_tot}
    \hat{H}_{\col}^{(2)}=\int_0^\infty \frac{\dd{E}}{2\pi} \int\frac{\dd[2]{\vec{k}}}{(2\pi)^2}\int_0^{2\pi}\frac{d\phi}{2\pi} \; E \left[\mathcal{A}_\theta^\dagger(E,\vec{k},\phi)\mathcal{A}_\theta(E,\vec{k},\phi)-\tilde{\mathcal{A}}_\theta^\dagger(E,\vec{k},\phi)\tilde{\mathcal{A}}_\theta(E,\vec{k},\phi)\right] \, .
\end{align}
Furthermore, let us denote $\mathcal{A}_L$ and $\mathcal{A}_R$ the inverse Bogoliubov transformation of $\mathcal{A}_{\theta}$ and $\tilde{\mathcal{A}}_{\theta}$ induced by $\hat{G}$:  
\begin{equation}
\label{eq:b_gff}
\begin{split}
    \mathcal{A}_L(E,\vec{k},\phi) = \ee^{\ii \hat{G}} \mathcal{A}_{\theta}(E,\vec{k},\phi) \ee^{- \ii \hat{G}} \, ,\\
    \mathcal{A}_R(E,\vec{k},\phi) = \ee^{\ii \hat{G}} \tilde{\mathcal{A}}_{\theta}(E,\vec{k},\phi) \ee^{- \ii \hat{G}} \, ,
\end{split}
\end{equation}
with
\begin{equation}
\tanh{\theta(E)} = e^{-\beta E/2} \, .
\end{equation}
They satisfy the canonical commutation relations
\begin{equation}
\label{eq:comm}
   [\mathcal{A}_L(E,\vec{k},\phi),\mathcal{A}_L^\dagger(E',\vec{k}',\phi')]=[\mathcal{A}_R(E,\vec{k},\phi),\mathcal{A}_R^\dagger(E',\vec{k}',\phi')]=\delta(E-E')\delta^{(2)}(\vec{k}-\vec{k}')\delta(\phi-\phi') \, .
\end{equation}
As before, these operators annihilate the ground state $\ket{0}$ instead of the thermal state $\ket{0(\beta)}$. Their thermal expectation values are 
\begin{align}
    \braket{\mathcal{A}_L(E,\vec{k},\phi)\mathcal{A}_L^\dagger(E',\vec{k}',\phi')}_\theta=&\cosh^2\theta(E)\delta(E-E')\delta^{(2)}(\vec{k}-\vec{k}')\delta(\phi-\phi') \, ,\\
    \braket{\mathcal{A}_L^\dagger(E,\vec{k},\phi)\mathcal{A}_L(E',\vec{k}',\phi')}_\theta=&\sinh^2\theta(E)\delta(E-E')\delta^{(2)}(\vec{k}-\vec{k}')\delta(\phi-\phi') \, ,
\end{align}
and similarly for $\mathcal{A}_R$.

\paragraph{}
For $\hat{G}$ we can expand it in the $1/N$:
\begin{equation}
    \hat{G} = \sqrt{N} \hat{G}^{(1)} + \hat{G}^{(2)} + \dots \, .
\end{equation}
Here $\hat{G}^{(1)}$ is linear in the bi-local fields, and will be eliminated by collective coordinate method in Section \ref{sec:colcoord}. $\hat{G}^{(2)}$ can be expressed in terms of bulk fields as
\begin{equation}
\label{eq:G_gff}
     \hat{G}^{(2)} =  \ii\int_0^\infty \frac{\dd{E}}{2\pi} \int\frac{\dd[2]{\vec{k}}}{(2\pi)^2}\int_0^{2\pi}\frac{d\phi}{2\pi}~ \theta(E)  \Big[\mathcal{A}_\theta^{\dagger}(E,\vec{k},\phi) \widetilde{\mathcal{A}}_\theta^{\dagger}(E, \vec{k}, \phi) - \mathcal{A}_\theta(E, \vec{k} ,\phi) \widetilde{\mathcal{A}}_\theta(E, \vec{k}, \phi)\Big] \, ,
\end{equation}
such that it gives the correct Bogoliubov transformation on $\mathcal{A}_\theta$
\begin{equation}
   \mathcal{A}_L(E, \vec{k} ,\phi) = \cosh{\theta(E)} \mathcal{A}_\theta (E, \vec{k} ,\phi) + \sinh{\theta(E)} \Tilde{\mathcal{A}}_\theta^\dagger (E, \vec{k} ,\phi) \, ,
\end{equation}
and similarly for $\mathcal{A}_R$.
Therefore at the quadratic order, in \eqref{eq:H_tot} and \eqref{eq:G_gff} $\mathcal{A}_\theta$ and $\Tilde{\mathcal{A}}_\theta$ can be replaced by $\mathcal{A}_L$ and $\mathcal{A}_R$
\begin{align}
    \hat{H}_{\col}^{(2)}=&\int_0^\infty \frac{\dd[2]{E}}{2\pi} \int\frac{\dd{\vec{k}}}{(2\pi)^2}\int_0^{2\pi}\frac{d\phi}{2\pi} \; E \left[\mathcal{A}_L^\dagger(E,\vec{k},\phi)\mathcal{A}_L(E,\vec{k},\phi)-\mathcal{A}_R^\dagger(E,\vec{k},\phi)\mathcal{A}_R(E,\vec{k},\phi)\right] \, , \\
    \hat{G}^{(2)} =& \ii\int_0^\infty \frac{\dd{E}}{2\pi} \int\frac{\dd[2]{\vec{k}}}{(2\pi)^2}\int_0^{2\pi}\frac{d\phi}{2\pi}~ \theta(E)  \Big[\mathcal{A}_L^{\dagger}(E,\vec{k},\phi) \mathcal{A}_R^{\dagger}(E, \vec{k}, \phi) - \mathcal{A}_L(E, \vec{k} ,\phi) \mathcal{A}_R(E, \vec{k}, \phi)\Big] \, .
\end{align}

\paragraph{}
To summarize, from collective fluctuations we see a complete spectrum of left and right bulk free fields $\mathcal{A}_L$ and $\mathcal{A}_R$ satisfying commutation relations \eqref{eq:comm}. These are extensive, with $0\leq E<\infty$, and reflect bulk spectra in the presence of a horizon. The Hilbert space is a product of left and the right commuting sub-algebras
\begin{equation}
    \{\mathcal{A}_L\}\otimes\{\mathcal{A}_R\} \, .
\end{equation}
We have also seen that the $\hat{G}$-symmetry has re-emerged at the level of quadratic fluctuations, taking the form of bulk level Bogoliubov transformations. Note also that the zero modes at  ($E=0$) are actually not present in the Hamiltonian $H^{(2)}_{\col}$. These will be replaced by collective coordinates in the full treatment in \Cref{sec:colcoord}.

\section{\texorpdfstring{$1/N$}{1/N} Expansions and Collective Coordinates}
\label{sec:colcoord}

Quantization around the thermal and extend soliton case share similarities. In both cases quadratic fluctuations are characterized by zero modes which are related to the associated symmetries. In the TFD we have two operators $H_{+}$ and $\hat{G}$ that commute with the Hamiltonian $\hat{H}$. These symmetry operators respectively give two classes of zero modes, $u_{l}$ and $v_{l}$ (to be revealed below). These operators when expanded at large $N$ again start with a linear term of order $\sqrt{N}$. (The leading term of $H_{+}$ is a c-number of order $N$, representing the thermal energy, which we subtract.) These `large operators' therefore also present problems in the naive $1/N$ expansions: the symmetry transformations are only implemented after an infinite re-summation. We will implement them through collective coordinates. There is a difference, however, between the two symmetries $H_{+}$ and $\hat{G}$, with the later acting through a unitary transformation. In particular, the counterpart for the center of mass of the soliton is the hyperbolic angle $\theta(\vec{k})$, and the state $\ket{s, x = 0}$ corresponds to the zero temperature state $\ket{0}$. Also, the analog of the zero modes $f_0$ is $v_{l}$ (zero modes of the potential matrix $V$) in TFD. To implement the collective coordinate method, we should impose the gauge condition
\begin{equation}
    \hat{g}(\vec k) - \hat{\mathcal{G}}(\vec k) \ket{0(\beta)} = 0 \, .
\end{equation}
The gauge condition can be chosen arbitrarily, and we will require
\begin{equation}
    \mathcal{H}_{+, - \hat{q}}(\vec k) \ket{0(\beta)} = 0 \, ,
\end{equation}
with $\hat{q}$ the canonical conjugate of $\hat{g}$. 

\subsection{Zero modes and large operators}

Let us first briefly discuss the relation between symmetry operators and zero modes. Here we will be schematic, and the complete expressions of the formulae will be presented in detail in the following subsections. Consider the quadratic Hamiltonian that we established in the previous section:
\begin{equation*}
    \hat{H}^{(2)}_{\col} = \frac{1}{2}\operatorname{Tr}[\boldsymbol{\pi}^{\operatorname{T}} K \boldsymbol{\pi} + \boldsymbol{\eta}^{\operatorname{T}} V \boldsymbol{\eta}] \, .
\end{equation*}
The kinetic matrix $K$ \eqref{eq:K} and the potential matrix $V$ \eqref{eq:V} were both seen to posses zero modes \cite{Jevicki:2021ddf}:
\begin{align}
    \Tr(K u_k)  = 0 \, ,
\end{align}
and
\begin{align}
    \Tr(V v_k) = 0 \, ,
\end{align}
with $k \equiv |\vec k|$ representing a label of the zero modes. These zero modes are associated with the order $\sqrt{N}$ terms of the symmetry operators. We have\footnote{
    Here inside the traces one has one bi-local field and some $c$-numbers, and the notation means $\Tr[c_1 u^{\operatorname{T}} \boldsymbol{\eta}] \equiv \int \frac{\dd[d]{\vec k}}{(2\pi)^d} c_1(\vec k) u_k \boldsymbol{\eta}(\vec k, \vec k)$, and similarly in the following discussions.
}
\begin{align}
    H_{+}^{(1)} & = \Tr[c_1 u^{\operatorname{T}} \boldsymbol{\eta}] \, , \\
    \hat{G}^{(1)} & = \Tr[c_2 v^{\operatorname{T}} \boldsymbol{\pi}] \, ,
\end{align}
with $c_1$ and $c_2$ some functions of $|\vec k|$ (see following for their precise expressions). We will see that these zero modes arise from the symmetry conditions when expanded order by order in $1/N$. Especially, at order $\sqrt{N}$ we have
\begin{align}
    0 & = [H_{+}^{(1)}, \hat{H}_{\col}^{(2)}] = \Tr(c_1 \boldsymbol{\pi}^{\operatorname{T}} K u) \, , \\
    0 & = [\hat{G}^{(1)}, \hat{H}_{\col}^{(2)}] = \Tr(c_2 \boldsymbol{\eta}^{\operatorname{T}} V v) \, .
\end{align}
Detailed calculations of these will be presented in the following.

\paragraph{}
These results are not limited to vector models. In matrix models these zero modes will appear in order $N$ terms of the symmetry operators. In particular, for large $N$ matrix quantum mechanics, we will have one $u$ and one $v$, and
\begin{align}
    H_{+}^{(1)} & = c_1^{\prime} u^{\operatorname{T}} \boldsymbol{\eta} \, , \\
    \hat{G}^{(1)} & = c_2^{\prime} v^{\operatorname{T}} \boldsymbol{\pi} \, , 
\end{align}
with $\boldsymbol{\eta}$ denoting a complete set of single trace operators and $\boldsymbol{\pi}$ its canonical conjugate. We stress that the symmetries $H_{+}$ and $\hat{G}$ explain zero modes of the kinetic and potential term in the quadratic $\hat{H}_{\col}^{(2)}$. We also note that (in this vector case) the space (of zero modes) induced by $\hat{G}$ is larger then the space induced by $H_{+}$. In the quantum mechanical matrix model case these sets are the same as both generate a one parameter symmetry.

\subsection{\texorpdfstring{$H_{+}$}{H plus} }
\label{subsec:average_energy_operator}

We will first give the $1/N$ expansion properties of $H_{+}$ and also give its collective coordinate version. The full Hamiltonian $H_{+}$, when written in terms of collective fields reads:
\begin{align}
    H_{+, \col} = & \frac{2}{N} \Tr[\Pi \star \Phi \star \Pi] + \frac{N}{8} \Tr[\Phi^{-1}] + \frac{N}{2} \Tr[-\nabla^2 \Phi|_{\vec x = \vec y}] \, .
\end{align}
Expanding it around the thermal background generates a $1/N$ expansion series
\begin{align}
    H_{+, \col}[\boldsymbol{\pi}, \boldsymbol{\eta}] = & \sum_{n=0}^{\infty} N^{1 - \frac{n}{2}} H_{+, \col}^{(n)}[\boldsymbol{\pi}, \boldsymbol{\eta}] \nonumber \\
    = & N H_{+, \col}^{(0)} + \sqrt{N} H_{+, \col}^{(1)} + H_{+, \col}^{(2)} + \dots \, .
\end{align}
The leading order (i.e. $n=0$) however is non-zero, and gives (twice of) the average energy $E(\beta)$ at inverse temperature $\beta$:
\begin{equation}
    E(\beta) = N H_{+, \col}^{(0)} = N \Tr[\omega \cosh(2 \theta)] \, .
\end{equation}
At the same time, the sub-leading term (i.e. $n=1$) does not vanish, and is given by
\begin{equation} \label{eq:Hplus_col_1}
    \sqrt{N} H_{+, \col}^{(1)} = \sqrt{N} \Tr[\omega^2 \sinh(2 \theta) u^{\operatorname{T}} \boldsymbol{\eta}] \, .
\end{equation}
Within the trace we have
\begin{equation} \label{eq:zero_mode_u}
    u_{l}(\vec{k}_1, \vec{k}_2) = \delta(|\vec{k}_1| - l) \delta^{d}(\vec{k}_1 - \vec{k}_2)
    \begin{pmatrix}
        - \sinh 2 \theta(\vec{k}_1) \\
        \cosh 2 \theta(\vec{k}_1) \\
        \cosh 2 \theta(\vec{k}_1) \\
        - \sinh 2 \theta(\vec{k}_1)
    \end{pmatrix} \, ,
\end{equation}
and we are integrating $u_{l} \equiv \int u_{l}(\vec{k}_1, \vec{k}_2) \dd[d]{\vec{k}_1} \dd[d]{\vec{k}_2}$. They are zero modes of $K$ \cite{Jevicki:2015sla,Jevicki:2021ddf} given in \cref{eq:zero_mode_u}:
\begin{equation}
    \int K(\vec{k}_1, \vec{k}_2)\cdot u_{l}(\vec{k}_1, \vec{k}_2) \dd[d]{\vec{k}_1} \dd[d]{\vec{k}_2} = 0 \, ,    
\end{equation}
which appear as the consistency condition that $H_{+}$ and $\hat{H}$ must commute with each other:
\begin{equation}
    [H_{+,\col}, \hat{H}_{\col}] 
    = \sum_{n=3}^{\infty} \sum_{m=2}^{n-1} N^{2 - \frac{n}{2}} [H_{+,\col}^{(n-m)}, \hat{H}_{\col}^{(m)}]
    = 0 \, .
\end{equation}
In the large $N$ limit, this gives a set of consistency equations at each order of $1/N$:
\begin{equation}
    \sum_{m=2}^{n-1} [H_{+, \col}^{(n-m)} , \hat{H}_{\col}^{(m)}] = 0 \, , \qquad n = 3, 4, \dots .
\end{equation}
Taking $n = 3$ we have
\begin{equation}
    [H_{+, \col}^{(1)}, \hat{H}_{\col}^{(2)}] 
    = \ii \int \omega_{\theta}^2 \sinh[2 \theta(\vec{k})] \boldsymbol{\pi}^{\operatorname{T}}(\vec{k}, \vec{k}) K(\vec{k}, \vec{k}) u_{k} \frac{\dd[d]{\vec{k}}}{(2\pi)^{d}} = 0 \, ,
\end{equation}
which explicitly shows that $u$ must be the zero mode of $K$.

\paragraph{}
In principle we can treat $H_{+}$ as a collective coordinate. To apply the collective coordinate method, we demand the constraint equation to be
\begin{equation}
    \hat{h} - H_+ \ket{0(\beta)} = 0 \, .
\end{equation}
We consider the critical point for simplicity, such that we can use conformal symmetry. Then we can choose the canonical gauge condition
\begin{equation}
\label{eq:GC_h}
    \hat{q} - \frac{D_+}{H_+} \ket{0(\beta)} = 0 \, ,
\end{equation}
where $D_+ = D_1 + D_2$ is the dilatation operator, such that
\begin{equation}
    [\hat{h}, \hat{q}] = \ii \, .
\end{equation}
One can shift the appearance of $\beta$ to $\beta + i\hat{q}$ by redefining the states and operators in the following way
\begin{equation}
    O'(\hat{q}) = e^{i \hat{q} H_+} O 
    \ee^{-\ii \hat{q} H_+} \, ,
\end{equation}
so that the oscillators are factorized into three commuting sets
\begin{equation}
    \{A_L\} \otimes \{\hat{q},\hat{h}\} \otimes \{A_R\} \, .
\end{equation}

\subsection{\texorpdfstring{$\hat{G}$}{G} as a collective coordinate}

Let us then demonstrate the implementation of the collective coordinate method with $\hat{G}$ in TFD. First we reveal the zero mode problems and the issues of $\hat{G}$-transformation at large $N$. For simplicity we will again consider the free theory case. In this case the potential matrix $V$ \eqref{eq:V0} have the zero modes \cite{Jevicki:2021ddf} (we omit the normalization factors):
\begin{equation} \label{eq:zero_mode_v}
    v_{l}(\vec{k}_1, \vec{k}_2) = \delta(|\vec{k}_1| - l) 
    \delta^{d}(\vec{k}_1 - \vec{k}_2)
    \begin{pmatrix}
        \sinh 2 \theta(\vec{k}_1) \\
        \cosh 2 \theta(\vec{k}_1) \\
        \cosh 2 \theta(\vec{k}_1) \\
        \sinh 2 \theta(\vec{k}_1)
    \end{pmatrix} \, .
\end{equation}
They obey
\begin{equation}
    \int V(\vec{k}_1, \vec{k}_2) v_{l}(\vec{k}_1, \vec{k}_2) \dd[d]{\vec{k}_1} \dd[d]{\vec{k}_2} = 0 \, . 
\end{equation}
The appearance of these modes has been understood well in \cite{Jevicki:2021ddf}: they are the Goldstone modes, and thus can be computed directly via 
\begin{equation}
    v_{l}(\vec{k}_1, \vec{k}_2) = \fdv{\Phi_{\theta}(\vec{k}_1, \vec{k}_2)}{\theta(\vec{l})} \, ,
\end{equation}
up to some coefficients, with $\Phi_{\theta}(\vec{k}_1, \vec{k}_2)$ the Fourier transform of the thermal background solution. We note that $u_{l}(\vec{k}_1, \vec{k}_2)$ and $v_{q}(\vec{k}_1, \vec{k}_2)$ obey the orthogonality condition
\begin{equation}
    \int u_{l}(\vec{k}_1, \vec{k}_3)^{\operatorname{T}} v_{q}(\vec{k}_3, \vec{k}_2) \dd[d]{\vec{k}_3} = 2 \delta(l - q) \delta^{d}(\vec{k}_1 - \vec{k}_2) \, .
\end{equation}
This relation implies that a class of degrees of freedom is completely missing in $\hat{H}$, as we have seen above. In addition, it also reveals that $H_{+}$ and $\hat{G}$ are not unrelated.

\paragraph{}
In the soliton problem, the zero mode $f_0$ can be derived from the symmetry condition $[H, P] = 0$. We can also show that the zero modes $v_{l}$ can be derived from $[\hat{\mathcal{G}}(\Vec{k}), \hat{H}] = 0$. Let us give a brief demonstration, again using the  free theory to simplify calculations. Since $\hat{\mathcal{G}}(\Vec{k})$ itself is invariant under the Bogoliubov transformation, we can write it as (cf. \cref{eq:G2})
\begin{align}
    \hat{\mathcal{G}}(\vec{k}) & = \sqrt{N} \hat{\mathcal{G}}^{(1)}(\vec{k}) + O(1) \\
    & =  \frac{2 \sqrt{N}}{\omega} \pi_{\theta}^{12}(\vec{k}, \vec{k}) + O(1) \\
    & =  \frac{2 \sqrt{N}}{\omega} v_{k}^{\operatorname{T}} \boldsymbol{\pi}(\vec{k}, \vec{k}) + O(1) \, .
\end{align}
In the last step, we use Bogoliubov transformation to write $\pi_{\theta}^{12}(\vec{k}, \vec{k}) = v_{k}^{\operatorname{T}}\boldsymbol{\pi}(\vec{k}, \vec{k})$. Here $v_{k} \equiv \int v_{k}(\vec{k}_1, \vec{k}_2) \dd[d]{\vec{k}_1} \dd[d]{\vec{k}_2}$, with the integrand the zero mode of $V(\vec{k}_1, \vec{k}_2)$. Then we compute the commutator to obtain
\begin{equation}
    0 = [\hat{H}_{\col}^{(2)}, \hat{\mathcal{G}}(\vec{k})] =  \frac{2\ii \sqrt{N}}{\omega} \boldsymbol{\eta}^{\operatorname{T}}(\vec{k}, \vec{k}) V(\vec{k}, \vec{k}) v_{k} + O(1) \, ,
\end{equation}
which indicates that $v$ must be zero modes of $V$.

\paragraph{}
Due to the dependence on $N$, issues arise when performing $\hat{G}$-transformations. We perform a $1/N$ expansion for $\hat{G}$:
\begin{equation}
    \hat{G} = \sqrt{N} \hat{G}^{(1)} + \hat{G}^{(2)} \, , 
\end{equation}
with
\begin{equation}
    \hat{G}^{(1)} = \ii \int \theta(\Vec{k})\,\hat{\mathcal{G}}^{(1)}(\vec{k}) \dd[d]{\vec{k}} \, .
\end{equation}
Due to the appearance of this order $\sqrt{N}$ term in $\hat{G}$, we have similar problem when transforming fields as in the soliton case. Again, let us illustrate this in an analogous pattern. We consider the following transformation of the bi-local field $\Phi(\vec{x}_1, \vec{x}_2)$, before which each field is expanded in $1/N$:
\begin{align}
    \label{eq:G_transformation_issue_illustration_1}
    \ee^{- \ii \hat{G}} \Phi(\vec{x}_1, \vec{x}_2) \ee^{\ii \hat{G}} & = 
        \ee^{- \ii \sqrt{N} \hat{G}^{(1)} - \ii \hat{G}^{(2)} + \cdots} (
            \Phi_{0} + N^{-1/2} \boldsymbol{\eta}
        ) 
        \ee^{\ii \sqrt{N} \hat{G}^{(1)} + \ii \hat{G}^{(2)} + \cdots} \\
        \label{eq:G_transformation_issue_illustration_2}
        & = \sum_{n=0}^{\infty} \frac{(- \ii)^n}{n!} \ad_{\sqrt{N} \hat{G}^{(1)} + \hat{G}^{(2)} + \cdots}^{n}(
            \Phi + N^{-1/2} \boldsymbol{\eta}
            ) \\
        \label{eq:G_transformation_issue_illustration_3}
        & = \Phi_{\theta}(\vec{x}_1, \vec{x}_2) + N^{-1/2} \boldsymbol{\eta}_{\theta}(\vec{x}_1, \vec{x}_2) \, ,
\end{align}
Here $\ad_{A}(B) = [A, B]$. In the first step \cref{eq:G_transformation_issue_illustration_1} we expand both $\hat{G}$ and $\Phi$ in $1/N$. Note that $\Phi_{0}$ is the large $N$ background at zero temperature. We expect the results would be \cref{eq:G_transformation_issue_illustration_3}. However, we see from \cref{eq:G_transformation_issue_illustration_2} that this computation cannot be done unless we know the full series of the $1/N$ expansions: different orders in $1/N$ get mixed. For example, consider the commutator $[\sqrt{N} \hat{G}^{(1)}, N^{-1/2} \eta]$, we see that the result gives an order 1 term, and hence contributes to the background, and shifts $\Phi_{0}$. 

\paragraph{}
Next,to decouple variables (in the gauge condition) we make a change of reference frame. In O($N$) vector TFD, we have the $\hat{G}$ operator as
\begin{equation}
    \hat{G} = \int \theta(\vec{k}) (\phi^i(\vec{k}) \tilde{\pi}^{i}(\vec{k}) + \tilde{\phi}^{i}(\vec{k})\pi^{i}(\vec{k})) \dd[d]{\vec{k}_1} \dd[d]{\vec{k}_2} \, .
\end{equation}
Define
\begin{equation}
    \Delta(\vec{k}_1, \vec{k}_2) = \begin{pmatrix}
        0 & \theta(\vec{k}_1) & \theta(\vec{k}_2) & 0 \\
        \theta(\vec{k}_1) & 0 & 0 & \theta(\vec{k}_2) \\
        \theta(\vec{k}_2) & 0 & 0 & \theta(\vec{k}_1) \\
        0 & \theta(\vec{k}_2) & \theta(\vec{k}_1) & 0 
    \end{pmatrix} \, .
\end{equation}
Then the bi-local collective field representation of $\hat{G}$ is given by
\begin{equation}
    \hat{G} = \Tr(\Pi^{\operatorname{T}} \, \Delta \, \Phi) \, ,
\end{equation}
where 
\begin{equation}
    \Tr(\Pi^{\rm T} \Delta \Phi) \equiv \int \dd[d]{\vec{k}_1} \dd[d]{\vec{k}_2} 
    \Pi^{\rm{T}}(\vec{k}_1, \vec{k}_2) \Delta(\vec{k}_1, \vec{k}_2) \Phi(\vec{k}_1, \vec{k}_2) \, .
\end{equation}
To introduce the collective coordinate, we need to extract the function $\theta(\vec{k})$. This can be done by introduce
\begin{equation}
    Q_{\vec{l}}(\vec{k}_1, \vec{k}_2) = 
    \begin{pmatrix}
        0 & \delta_{\vec{k}_1, \vec{l}} & \delta_{\vec{k}_2, \vec{l}} & 0 \\
        \delta_{\vec{k}_1, \vec{l}} & 0 & 0 & \delta_{\vec{k}_2, \vec{l}} \\
        \delta_{\vec{k}_2, \vec{l}} & 0 & 0 & \delta_{\vec{k}_1, \vec{l}} \\
        0 & \delta_{\vec{k}_2, \vec{l}} & \delta_{\vec{k}_1, \vec{l}} & 0
    \end{pmatrix} \, ,
\end{equation}
such that
\begin{equation}
    \Delta(\vec{k}_1, \vec{k}_2) = \int \,  \dd[d]{\vec{l}} \, \theta(\vec{l}) \, Q_{\vec{l}}(\vec{k}_1, \vec{k}_2) \, .
\end{equation}
The $\hat{G}$ operator then can be written as
\begin{equation}
    \hat{G} = \int \dd[d]{k} \theta(\vec{k}) \hat{\mathcal{G}}(\vec{k}) \, , 
    \qquad
    \hat{\mathcal{G}}(\vec{k}) = \Tr(\Pi^{T} \, Q_{\vec{k}} \, \Phi) \, .
\end{equation}
We then introduce the collective coordinate $\hat{q}(\vec{k})$ to define
\begin{equation}
    \hat{G}_{\hat{q}} = \int \dd[d]{\vec{k}} \hat{q}(\vec{k}) \hat{\mathcal{G}}(\vec{k}) \, .
\end{equation}
This is the analog of $\hat{x} P$ in the one-soliton case. There we only have one zero mode, and hence we have one collective coordinate $\hat{x}$. Here we have infinitely many zero modes, and thus we introduce $\hat{q}(\vec{k})$.  One can redefine the states and collective operators in the following way
\begin{equation}
\mathcal{O}' \equiv \mathcal{O}_{-\hat{q}} = \ee^{-\ii  \hat{G}_{\hat{q}} } 
\mathcal{O}
\ee^{\ii \hat{G}_{\hat{q}} }  \, .
\end{equation}
For bi-local field, we can write down the explicit transformation
\begin{equation}
\boldsymbol{\eta}'(\vec{k}_1, \vec{k}_2) = M[-\hat{q}(\vec{k}_1), -\hat{q}(\vec{k}_2)] \boldsymbol{\eta}(\vec{k}_1, \vec{k}_2) \, .
\end{equation}

\paragraph{}
We then introduce the constraint and the gauge condition
\begin{align}
    \hat{g}(\vec{k}) - \hat{\mathcal{G}}[\Pi, \Phi](\vec{k}) \ket{0(\beta)} & = 0 \, , \\
    \chi_{-\hat{q}}[\Pi, \Phi](\vec{k}) \ket{0(\beta)} & = 0 \, .
\end{align}
 Here $\hat{g}(\vec{k})$ is the canonical conjugate of $\hat{q}(\vec{k})$. The oscillators are factorized into three commuting sets
\begin{equation}
\{A_L\} \otimes \{ \hat{q}, \hat{g}\} \otimes \{A_R\} \, .
\end{equation}
The gauge condition is chosen to be of this form so that the $\hat{G}_{\hat{q}}$-transformation can be undone and the $\hat{q}$-dependence can be eliminated by switching to the centre frame of reference. To be more explicit, we make a transformation
\begin{align}
    \label{eq:Phi_prime}
    \Phi^{\prime}(\vec{k}_1, \vec{k}_2) & = 
    \ee^{-\ii \hat{G}_{\vec{q}}} \Phi(\vec{k}_1, \vec{k}_2) \ee^{\ii \hat{G}_{\vec{q}}} =
     \exp[-\int \dd[d]{\vec{k}} \hat{q}(\vec{k}) Q_{\vec{k}}(\vec{k}_1, \vec{k}_2)] \Phi(\vec{k}_1, \vec{k}_2) \, , \\
    \label{eq:Pi_prime}
    \Pi^{\prime}(\vec{k}_1, \vec{k}_2) & =
     \ee^{-\ii \hat{G}_{\vec{q}}} \Pi(\vec{k}_1, \vec{k}_2) \ee^{\ii \hat{G}_{\vec{q}}} = 
     \exp[\int \dd[d]{\vec{k}} \hat{q}(\vec{k}) Q_{\vec{k}}(\vec{k}_1, \vec{k}_2)] \Pi(\vec{k}_1, \vec{k}_2) \, .
\end{align}
Consequently, the constraint and the gauge condition become
\begin{align}
    \hat{g}(\vec{k}) - \hat{\mathcal{G}}[\Pi^{\prime}, \Phi^{\prime}](\vec{k}) \ket{0(\beta)}^{\prime} & = 0 \, , \\
    \chi[\Pi^{\prime}, \Phi^{\prime}](\vec{k}) \ket{0(\beta)}^{\prime} & = 0 \, .
\end{align}
We then perform a $1/N$ expansion of the fields
\begin{equation}
    \Phi^{\prime}  = \Phi_{\theta}^{\prime} + \frac{1}{\sqrt{N}} \boldsymbol{\eta}^{\prime} \, , 
    \qquad
    \Pi^{\prime} = \sqrt{N} (\boldsymbol{\pi}_{\theta}^{\prime} + \boldsymbol{\pi}^{\prime}) \, .
\end{equation}
One can check that the zero modes are given by
\begin{align}
    u_{\vec k}(\vec k_1, \vec k_2) & = Q_{\vec k}(\vec k_1, \vec k_2) \Phi_{-\theta}(\vec k_1, \vec k_2) \, , 
    \qquad \text{zero modes of } K \, , \\
    v_{\vec k}(\vec k_1, \vec k_2) & = Q_{\vec k}(\vec k_1, \vec k_2) \Phi_{\theta}(\vec k_1, \vec k_2) \, , 
    \qquad \text{zero modes of } V \, .
\end{align}
Using these relations, the constraint is expanded to be
\begin{equation}
    \hat{g}(\vec{k}) - \sqrt{N} \Tr(\boldsymbol{\pi}_{\theta}^{\prime} Q_{\vec{k}} \Phi^{\prime}) 
    - \sqrt{N} \Tr(\boldsymbol{\pi}^{\prime} v_{\vec k}) 
    - \Tr(\boldsymbol{\pi}^{\prime} Q_{\vec{k}} \boldsymbol{\eta}^{\prime}) \ket{0(\beta)}^{\prime} = 0 \, .
\end{equation}
We have the exact solution of $\boldsymbol{\pi}_\theta^{\prime}$:
\begin{equation}
    \boldsymbol{\pi}_{\theta}^{\prime}  = \frac{u_{\vec k}^{\prime}}{\sqrt{N}}
         \frac{
            \hat{g}(\vec{k}) - \Tr(\boldsymbol{\pi}^{\prime} Q_{\vec{k}} \boldsymbol{\eta}^{\prime})
         }{
        \Tr(u_{\vec k}^\prime Q_{\vec{k}} \Phi^{\prime})
         } \, .
\end{equation}
This solution can be expanded in $1/N$,
\begin{equation}
\boldsymbol{\pi}_{\theta}^{\prime} = \frac{u_{\vec k}^{\prime}}{2\sqrt{N}}\hat{g}(\vec{k}) - \frac{u_{\vec k}^{\prime}}{2\sqrt{N}}\Tr(\boldsymbol{\pi}^{\prime} Q_{\vec{k}} \boldsymbol{\eta}^{\prime}) - \frac{u_{\vec k}^{\prime}}{4N}\hat{g}(\vec{k})\Tr(u_{\vec k}^\prime Q_{\vec{k}} \boldsymbol{\eta}^{\prime}) + O(1) \, .
\end{equation}
At the same time, the zero modes $v$ are projected out from $\boldsymbol{\pi}^{\prime}$:
\begin{equation}
\label{eq:cons}
    \Tr(\boldsymbol{\pi}^{\prime} v_{\vec k}^\prime) \ket{0(\beta)}^{\prime} = 0 \, .
\end{equation}
The gauge condition $\chi$ can be arbitrary. We choose the simplest gauge
\begin{equation}
    \Tr(\boldsymbol{\eta}^{\prime} u_{\vec k}^\prime) \ket{0(\beta)}^{\prime}= 0 \, ,
\end{equation}
such that with \cref{eq:cons} the zero modes are projected out from the fluctuations (They are replaced by the collective coordinates $\{\hat{q}, \hat{g}\}$.). For these one has the subsidiary conditions:
\begin{equation}
    \hat{q}(\vec k) + \frac{\ii}{8 \omega(\vec{k}) N} \hat{g}(\vec k) \ket{0(\beta)} = 0 \, ,
\end{equation}
explained in \Cref{sec:cst_tfd}, which leads to the centre part of the wave functional
\begin{equation}
    \Psi_{\rm c}[q] = \mathcal{N_{\rm c}} \exp[- N \int 4 \omega(\vec k) q^2(\vec k) \dd[d] \vec k] .
\end{equation}
Therefore the total wave functional of the TFD state reads
\begin{align}
    \Psi[q,\boldsymbol{\eta}'] = & \mathcal{N}\exp[- N \int 4 \omega(\vec k) q^2(\vec k) \dd[d] \vec k] \nonumber \\ 
    & \times \exp[
        - \frac{1}{2} \int \boldsymbol{\eta}^{\prime\operatorname{T}}(\vec{k}_1, \vec{k}_2) \, G^{\prime-1}(\vec{k}_1, \vec{k}_2) \, \boldsymbol{\eta}^\prime(\vec{k}_1, \vec{k}_2) \frac{\dd[d]{\vec{k}_1}}{(2\pi)^d} \frac{\dd[d]{\vec{k}_2}}{(2\pi)^d}
    ] \, .
\end{align}
As a result, the thermofield Hamiltonian becomes
\begin{equation}
    \hat{H}_{\col}^{(2)} = \frac{1}{2} \Tr(\boldsymbol{\pi}^{\prime \operatorname{T}} K^\prime \boldsymbol{\pi}^\prime + \boldsymbol{\eta}^{\prime \operatorname{T}} V^\prime \boldsymbol{\eta}^\prime ) \, .
\end{equation}
Comparing with the translational case, the collective coordinates $\hat{g}$ and their canonical conjugates $\hat{q}$ are completely absent from the thermofield Hamiltonian, which is consistent with the fact that
\begin{equation*}
    [\hat{G}, \hat{H}] = 0 \, .
\end{equation*}
One now has that both $\hat{H}$ and $H_+$ are expanded in $1/N$
\begin{align}
    \hat{H} =& \hat{H}^{(2)} + \frac{1}{\sqrt{N}}\hat{H}^{(3)} + \cdots \, , \\
    H_+ =& N H_+^{(0)} + H_+^{(2)} + \frac{1}{\sqrt{N}} H_+^{(3)} + \cdots \, ,
\end{align}
with the large $\sqrt{N}\hat{H}_+^{(1)}$ being put to zero (due to the gauge condition). They together with the well-defined propagator allow for a systematic $1/N$ expansion. At the quadratic level (and therefore the Hilbert space) the structure is similar to the one predicted by Witten \cite{Witten:2021unn} on very general grounds. However, in regards to perturbation expansion issues have been brought up in \cite{Witten:2021unn} which might be solved in the present treatment.

\section{Conclusions}

We have in the present work addressed the structure of the Hilbert space and the $1/N$ expansion in perturbation around large $N$ extended states. Fluctuations around these states are singular, and are characterized by zero modes associated with broken symmetries. This is generally addressed by introduction of quantum mechanical collective coordinates with a Hilbert space containing these in addition to the fluctuating bulk fields. In the TFD state case the diagonal gauging of O($N$) was emphasized, which as it was seen appears appropriate for a two-sided ER bridge space-time. Even though in the vector theory that was used in this study the dual higher spin (HS) theory involves all spins, one still expects that in the thermal case the appropriate linearized fluctuations are to be in gravitational two-sided backgrounds with a horizon. Clearly this is to be understood from analogous study at the HS level. Also gauging in CFT one has a parallel in recent gravitational studies (in two-sided wormhole space-times)~\cite{Mathur:2012dxa,Marolf:2012xe,Jensen:2014lua} with diagonal implementation of constraint symmetries. Finally the structure of the large $N$ Hilbert space, and implementation of Goldstone symmetries that we have exhibited with explicit evaluations in large $N$ vector case applies more generally, in particular in matrix type models. This follows from the general structure of the Hamiltonian collective field theory.

\acknowledgments{
We would like to acknowledge useful discussions with Sumit Das, Robert de Mello Koch, Samir Mathur, Sanjaye Ramgoolam and Junggi Yoon on topics related to this work. This work is supported by the U.S. Department of Energy under contract DE-SC0010010.
}


\appendix

\section{TFD State Decomposition}
\label{appendix:TFD_state_decomposition}

We now turn to the discussion of the TFD state. To illustrate the TFD wave functional, it is more illustrative to work in the normal modes basis. We have already seen that $M$ diagonalizes the matrices $K$ \eqref{eq:K} and $V$ \eqref{eq:V} simultaneously. In addition, it also diagonalizes $G^{-1}$:
\begin{equation}
    M[\theta_1, \theta_2]^{\operatorname{T}} G^{-1}(\vec{k}_1, \vec{k}_2) M[\theta_1, \theta_2]
    = \omega_1 \omega_2 \mathbb{1}_{4} \, .
\end{equation} 
This simple relation reveals that the TFD wave function is diagonalized in terms of the normal modes:
\begin{equation} \label{eq:Psi_beta_eta_theta}
    \Psi_{\beta}[\boldsymbol{\eta_{\theta}}] = \mathcal{N} \exp[- \frac{1}{2} 
        \int \omega_1 \omega_2 \boldsymbol{\eta}_{\theta}^{\operatorname{T}} \boldsymbol{\eta}_{\theta} 
        \frac{\dd[d]{\vec{k}_1}}{(2\pi)^d} \frac{\dd[d]{\vec{k}_2}}{(2\pi)^{d}}] \, .
\end{equation}

\paragraph{}
In particular, in terms of the normal modes $\boldsymbol{\eta}_{\theta}$, we can decompose the TFD wave functional $\Psi_{\beta}[\boldsymbol{\eta}]$ into three pieces:
\begin{equation}
    \Psi_{\beta}[\boldsymbol{\eta}] = \Psi_{\beta}^{\rm{ns}}[\boldsymbol{\eta}] \, \Psi_{\beta}^{\rm{s}}[\boldsymbol{\eta}] \, \Psi_{\beta}^{\rm{c}}[\boldsymbol{\eta}] \, .
\end{equation}
$\Psi_{\beta}^{\rm{ns}}[\boldsymbol{\eta}]$ corresponds to the non-singular part of the thermofield Hamiltonian $\hat{H}_{\col, \rm{ns}}^{(2)}$ \eqref{eq:H_col_ns}:
\begin{align}
    \Psi_{\beta}^{\operatorname{ns}}[\boldsymbol{\eta}] & = \Psi_{\beta}^{\operatorname{ns}}[\boldsymbol{\eta}_{\theta, |\vec{k}_1| \neq |\vec{k}_2|}]
    \nonumber \\ 
    & = \mathcal{N}_{\rm{ns}} \exp[
        - \frac{1}{2} \int_{| \vec{k}_1 | \neq | \vec{k}_2 |} \omega_{\theta}(\vec{k}_1) \omega_{\theta}(\vec{k}_2) \, \boldsymbol{\eta}_{\theta}^{\operatorname{T}}(\vec{k}_1, \vec{k}_2) \boldsymbol{\eta}_{\theta}(\vec{k}_1, \vec{k}_2)
        \frac{\dd[d]{\vec{k}_1}}{(2\pi)^{d}} \frac{\dd[d]{\vec{k}_2}}{(2\pi)^{d}}
    ] \, .
\end{align}
Similarly $\Psi_{\beta}^{\rm{s}}[\boldsymbol{\eta}]$ corresponds to the singular part of the thermofield Hamiltonian $\hat{H}_{\col, \rm{s}}^{(2)}$ \eqref{eq:H_col_s}:
\begin{align}
    \Psi_{\beta}^{\rm{s}}[\boldsymbol{\eta}] & = \Psi_{\beta}^{\rm{s}}[\eta_{\theta, |\vec{k}_1| = |\vec{k}_2|}^{11}, \eta_{\theta, |\vec{k}_1| = |\vec{k}_2|}^{22}] \nonumber \\ 
    & = \mathcal{N}_{\rm{s}} \exp[
        - \frac{1}{2} \int_{| \vec{k}_1 | = | \vec{k}_2 |}
        \omega_{\theta}(\vec{k}_1) \omega_{\theta}(\vec{k}_2) \, \Bigl(
            [\eta_{\theta}^{11}(\vec{k}_1, \vec{k}_2)]^{2} + [\eta_{\theta}^{22}(\vec{k}_1, \vec{k}_2)]^2
        \Bigr)
        \frac{\dd[d]{\vec{k}_1}}{(2\pi)^{d}} \frac{\dd[d]{\vec{k}_1}}{(2\pi)^{d}}] \, .
\end{align}

\paragraph{}
Lastly, the central sector of the TFD wave functional $\Psi_{\beta}^{\rm{c}}[\boldsymbol{\eta}]$ corresponds to the central part, and cannot be determined from $\hat{H}_{\col}^{(2)}$. To obtain $\Psi_{\beta}^{\rm{c}}[\boldsymbol{\eta}]$, we may add a regulator the $\hat{H}_{\col}^{(2)}$ of the form (in the following we take $\eta_{0}(\vec{k}_1, \vec{k}_2) \equiv \eta_{\theta}^{12}(\vec{k}_1,\vec{k}_2)|_{|\vec{k}_1|=|\vec{k}_2|}$ and similarly for its canonical conjugate)
\begin{equation}
    \Delta \hat{H} = \frac{1}{2} \int [ m(\vec{k}_1)\pi_{0}^{2}(\vec{k}_1, \vec{k}_2) + \mu(\vec{k}_1) \eta_{0}^{2}(\vec{k}_1, \vec{k}_2) ]
    \frac{\dd[d]{\vec{k}_1}}{(2\pi)^{d}} \frac{\dd[d]{\vec{k}_2}}{(2\pi)^d} \, ,
\end{equation}
with the functions $m(\vec{k})$ and $\mu(\vec{k})$ only depending on the norm of the momentum, and obeying
\begin{equation}
    m(\vec{k}_1) \rightarrow 0  \, , \qquad \mu(\vec{k}_1) \rightarrow 0 \, , \qquad 
    \text{with } \sqrt{\frac{\mu(\vec{k}_1)}{m(\vec{k}_1)}} = 2 \omega_{\theta}(\vec{k}_1) \text{ fixed.}
\end{equation}
This regulator commutes with both $H_{+}$ and $\hat{G}$ (and also $\hat{\mathcal{G}}$), since we take the limit $m(\vec{k}_1) \rightarrow 0$. The thermofield Hamiltonian $\hat{H}_{\col}^{(2)} + \Delta \hat{H}$ then has information for this zero mode, from which we can easily obtain $\Psi_{\beta}^{\rm{c}}$:
\begin{align}
    \Psi_{\beta}^{\rm{c}}[\boldsymbol{\eta}] & = \Psi_{\beta}^{\rm{c}}[\eta_{\theta, |\vec{k}_1| = |\vec{k}_2|}^{12}] \nonumber \\
    \label{eq:Psi_c}
    & = \mathcal{N}_{\rm{c}} \exp[
        - \int_{|\vec{k}_1| = |\vec{k}_2|} G_{\operatorname{c}}(\vec{k}_1, \vec{k}_2)^{-1} \,
            [\eta_{\theta}^{12}(\vec{k}_1, \vec{k}_2)]^{2} 
            \frac{\dd[d]{\vec{k}_1}}{(2\pi)^{d}} \frac{\dd[d]{\vec{k}_2}}{(2\pi)^d}
        ] \, .
\end{align}
The normalization factors satisfy $\mathcal{N} = \mathcal{N}_{\rm{ns}} \, \mathcal{N}_{\rm{s}} \, \mathcal{N}_{\rm{c}}$. Another way to see this is that the TFD wave functional is related to the equal-time thermal two-point functions of bi-local fields, as indicated in \cref{eq:G_correlator_relation}. Using the non-vanishing two-point functions (cf. \cref{eq:G_correlator_relation})
\begin{align}
    G_{\operatorname{c}}(\vec{k}_1, \vec{k}_2) & = \int \frac{\dd[d]{\vec{k}_2}}{(2\pi)^{d}} \frac{\dd[d]{\vec{k}_4}}{(2\pi)^d} \, 4 \langle \eta^{12}_{\theta}(\vec{k}_1, \vec{k}_2) \eta^{12}_{\theta}(\vec{k}_3, \vec{k}_4) \rangle_{\beta} \nonumber \\ 
    & = \int \frac{\dd[d]{\vec{k}_2}}{(2\pi)^{d}} \frac{\dd[d]{\vec{k}_4}}{(2\pi)^d} \frac{1}{\omega_{\theta}(\vec{k}_1) \omega_{\theta}(\vec{k}_2)} (2\pi)^d \delta^{d}(\vec{k}_1 - \vec{k}_3) (2\pi)^d \delta^{d}(\vec{k}_2 - \vec{k}_4) \nonumber \\
    & = \frac{1}{\omega_{\theta}(\vec{k}_1) \omega_{\theta}(\vec{k}_2)} \, ,
\end{align}
and taking $\eta_{\theta}^{21}(\vec{k}_1, \vec{k}_2)$ into account gives four identical terms, which cancels out the the numerical factor above, we immediately recover the central part of the wave function $\Psi_{\beta}^{\rm{c}}[\boldsymbol{\eta}]$. 

\paragraph{Note.} Concerning the central part of the wave function $\Psi_{\beta}^{\rm{c}}[\boldsymbol{\eta}]$, 
we denote $k \equiv |\vec{k}|$ and introduce the density fields for $H_{+,\col}^{(0)}$ and $W \equiv H_{+,\col}^{(1)}$ respectively as
\begin{align}
    \mathcal{H}_+^{(0)}(k) & = \omega_{\theta}(k) \cosh(2 \theta(k)) \, , \\
    \mathcal{W}(k) & = - 2 \int \omega_{\theta}^2(\vec{k}) \sinh(2 \theta(\vec{k})) \eta_{\theta}^{12}(\vec{k}, \vec{k}) \frac{\dd[d-1]{\Omega_{\vec{k}}}}{(2\pi)^d} \, ,
\end{align}
where $\Omega_{\vec{k}}$ denotes the solid angle variables in momentum space.
We write it as
\begin{equation}
    \Psi_{\beta}^{\rm{c}}[\boldsymbol{\eta}] = \mathcal{N}_{\rm{c}} \exp[
        - \frac{1}{8} \int_{0}^{\infty} \frac{\mathcal{W}^2(k)}{\abs{\partial_{\beta} \mathcal{H}_+^{(0)}(k)}}
    \dd{k} 
    ] \, .
\end{equation}
We see the analogy to Witten's result \cite{Witten:2021unn}, written in the form
\begin{equation}
    g(W) = \mathcal{N}_{g} \exp[- \frac{W^2}{4 \abs{\partial_{\beta} H_{+, \col}^{(0)}}}] \, .
\end{equation}

\section{Symmetry Transformations}
\label{appendix:Bogoliubov_transformations}

We summarize various Bogoliubov transformation of the thermofield dynamics in free O($N$) vector model. In TFD formalism, for an operator $\mathcal{O}$, its Bogoliubov transformation is given through the $\hat{G}$ operator by
\begin{equation}
    \mathcal{O}_{\theta} := \ee^{- \ii \hat{G}} \mathcal{O} \ee^{\ii \hat{G}} \, ,
\end{equation}
which satisfies
\begin{equation}
    \langle 0, \tilde{0} | \mathcal{O} | 0,  \tilde{0} \rangle = \bra{0(\beta)} \mathcal{O}_{\theta} \ket{0(\beta)} \, .
\end{equation}
Thus, the $\hat{G}$-transformation preserves symplectic structures. Recall that in free theory $\hat{G}^{(2)}$ \eqref{eq:G2} is given by
\begin{equation}
    \hat{G}^{(2)} = \Intk \ii \theta(\vec{k}) \Big(a^{\dagger i}(\vec{k}) \widetilde{a}^{\dagger i}(\vec{k}) - a^{i}(\vec{k}) \widetilde{a}^{i}(\vec{k})\Big) \, .
\end{equation}

\paragraph{\texorpdfstring{$\operatorname{O}(N)$}{O(N)} vector fields.}

Let $\xi^i(\vec{k}) = (a^i, \widetilde{a}^i, a^{\dagger i}, \widetilde{a}^{\dagger i})(\vec{k})$, by a direct calculation one can show
\begin{equation} \label{eq:Bogoliubov_transformation}
    \xi^{i}_{\theta}(\vec{k}) = 
    \ee^{- \ii \hat{G}} \xi^{i}(\vec{k}) \ee^{\ii \hat{G}}
    = U[- \theta(\vec{k})] \xi^{i}(\vec{k}) \, ,
\end{equation}
with the Bogoliubov transformation matrix $U$ as
\begin{equation}
    U[- \theta(\vec{k})] = 
    \begin{pmatrix}
        \cosh \theta(\vec{k}) & 0 & 0 & - \sinh \theta(\vec{k}) \\
        0 & \cosh \theta(\vec{k}) & - \sinh \theta(\vec{k}) & 0 \\
        0 & - \sinh \theta(\vec{k}) & \cosh \theta(\vec{k}) & 0 \\
        - \sinh \theta(\vec{k}) & 0 & 0 & \cosh \theta(\vec{k}) 
    \end{pmatrix} \, .
\end{equation}
The $U$ matrices obey $U[\theta(\vec{k})] U[\theta(\vec{p})] = U[\theta(\vec{k}) + \theta(\vec{p})] $, such that they form a one-(functional)-parameter group. Furthermore, we have $U \in \operatorname{Sp}(4, \mathbb{R})$, implying that it induces a canonical transformation.

\paragraph{}
We can also study the Bogoliubov transformations of the fields $\varphi^{i}$, $\widetilde{\varphi}^{i}$ and their canonical conjugates. Let $\chi^{i}(\vec{k}) = (\varphi^i, \widetilde{\varphi}^i, \pi^i, \widetilde{\pi}^i)(\vec{k})$, the $\hat{G}$-transformation can be written as
\begin{equation}
    \chi^{i}_{\theta}(\vec{k}) = \ee^{- \ii \hat{G}} \chi^{i}(\vec{k}) \ee^{\ii \hat{G}} 
    = U^{\prime}[- \theta(\vec{k})] \chi^{i}(\vec{k}) \, .
\end{equation}
The new Bogoliubov transformation matrix $U^{\prime}$ is now block-diagonal:
\begin{equation} \label{eq:Uprime_and_S}
    U^{\prime}[- \theta(\vec{k})] = \begin{pmatrix}
        S[- \theta(\vec{k})] & 0 \\
        0 & S[\theta(\vec{k})]
    \end{pmatrix} \, ,
    \quad
    S[\theta(\vec{k})] = \begin{pmatrix}
        \cosh \theta(\vec{k}) & \sinh \theta(\vec{k}) \\
        \sinh \theta(\vec{k}) & \cosh \theta(\vec{k})
    \end{pmatrix} \, .
\end{equation}
One can check that $U^{\prime}$ still obeys the properties listed above. At the same time, The $S$ matrices also form a one-(functional)-parameter group, and $S[\theta(\vec{k})] \in \operatorname{SU}(1,1)$.

\paragraph{Bi-local collective fields.}
Let us consider the Bogoliubov transformations induced by $\hat{G}$ of the bi-local fields $\eta$:
\begin{equation}
    \boldsymbol{\eta}_{\theta}(\vec{k}_1, \vec{k}_2) = \ee^{-\ii \hat{G}} \boldsymbol{\eta}(\vec{k}_1, \vec{k}_2) \ee^{\ii \hat{G}} 
\end{equation}
We find ($\theta_{a} \equiv \theta(\vec{k}_{a})$)
\begin{equation}
    \boldsymbol{\eta}_{\theta}(\vec{k}_1, \vec{k}_2) = M[-\theta_1, - \theta_2] \boldsymbol{\eta}(\vec{k}_1, \vec{k}_2) \, .
\end{equation}
Let
\begin{equation}
    \mathfrak{c}_i \equiv \cosh \theta(\vec{k}_i) \, , 
    \qquad
    \mathfrak{s}_i \equiv \sinh \theta(\vec{k}_i) \, ,
\end{equation}
$M[\theta(\vec{k}_1), \theta(\vec{k}_2)]$ is the tensor product of $S[\theta(\vec{k}_1)]$ and $S[\theta(\vec{k}_2)]$ \cref{eq:Uprime_and_S}, and can be written as:
\begin{equation}
    M[\theta_1, \theta_2] = S[\theta_1] \otimes S[\theta_2] =
    \begin{pmatrix}
        \mathfrak{c}_1 \mathfrak{c}_2 & \mathfrak{c}_1 \mathfrak{s}_2 &  \mathfrak{c}_2 \mathfrak{s}_1 &  \mathfrak{s}_1 \mathfrak{s}_2 \\
        \mathfrak{c}_1 \mathfrak{s}_2 &  \mathfrak{c}_1 \mathfrak{c}_2 &  \mathfrak{s}_1 \mathfrak{s}_2 &  \mathfrak{c}_2 \mathfrak{s}_1 \\
        \mathfrak{c}_2 \mathfrak{s}_1 &  \mathfrak{s}_1 \mathfrak{s}_2 &  \mathfrak{c}_1 \mathfrak{c}_2 &  \mathfrak{c}_1 \mathfrak{s}_2 \\
        \mathfrak{s}_1 \mathfrak{s}_2 &  \mathfrak{c}_2 \mathfrak{s}_1 &  \mathfrak{c}_1 \mathfrak{s}_2 &  \mathfrak{c}_1 \mathfrak{c}_2
    \end{pmatrix}
\end{equation}
We have two important properties of $M$:
\begin{enumerate}
    \item $M$ is a two-(functional)-parameter group:
    \begin{equation}
        M[\theta_1, \theta_2] M[\theta_3, \theta_4] = M[\theta_1 + \theta_3, \theta_2 + \theta_4] \, ,
    \end{equation}
    so that its inverse is 
    \begin{equation}
        M[- \theta_1, - \theta_2] = ( M[\theta_1, \theta_2] )^{-1} \, .
    \end{equation}

    \item $M \in \operatorname{SU}(2,2)$. Let $D = \operatorname{diag}\{1, -1, -1, 1\}$, then we have
    \begin{equation}
        M[\theta_1, \theta_2] D
        M[\theta_1, \theta_2]^{\operatorname{T}}
       = D \, .
    \end{equation}
\end{enumerate}

\paragraph{Collective oscillators.}
It is convenient to define matrices of the bi-local operators as in \cite{Jevicki:2015sla}:
\begin{equation}
    \mathbb{A}_{\theta}(\vec{k}_1, \vec{k}_2) = 
    \begin{pmatrix}
        A_{\theta}(\vec{k}_1, \vec{k}_2) & C_{\theta}(\vec{k}_1, \vec{k}_2) \\
        C_{\theta}(\vec{k}_2, \vec{k}_1) & \widetilde{A}_{\theta}(\vec{k}_1, \vec{k}_2)
    \end{pmatrix} \, , \quad
    \mathbb{A}_{\theta}^{\dagger}(\vec{k}_1, \vec{k}_2) = 
    \begin{pmatrix}
        A_{\theta}^{\dagger}(\vec{k}_1, \vec{k}_2) & C^{\dagger}_{\theta}(\vec{k}_1, \vec{k}_2) \\
        C_{\theta}^{\dagger}(\vec{k}_2, \vec{k}_1) & \widetilde{A}_{\theta}^{\dagger}(\vec{k}_1, \vec{k}_2) 
    \end{pmatrix} \, , 
\end{equation}
and
\begin{equation}
    \mathbb{B}_{\theta}(\vec{k}_1, \vec{k}_2) = 
    \begin{pmatrix}
        B_{\theta}(\vec{k}_1, \vec{k}_2) & D_{\theta}^{\dagger}(\vec{k}_1, \vec{k}_2) \\
        D_{\theta}(\vec{k}_2, \vec{k}_1) & \widetilde{B}_{\theta}(\vec{k}_1, \vec{k}_2) 
    \end{pmatrix} \, .
\end{equation}
The definition and the algebra of these composite operators $A$, $B$, $C$ and $D$ are summarized in Appendix A of \cite{Jevicki:2021ddf}. Here we will discuss their Bogoliubov transformations, and also the counterpart for their $1/N$ expansions.

\paragraph{}
In the large $N$ limit, these operator have $1/N$ expansions. To linear order we have\footnote{
     Here to emphasize the Bogoliubov transformation of fields, we put a subscript $\theta$ for the fields. Comparing with the notations in \cite{Jevicki:2015sla}, we have $\alpha_{\operatorname{there}} \equiv \alpha_{\theta, \operatorname{here}}$, and similarly for other bi-local oscillators.
 }
 \begin{align}
     \mathbb{A}_{\theta} = \sqrt{2 N} \begin{pmatrix}
         \alpha_{\theta} & \gamma_{\theta} \\
         \widetilde{\gamma}_{\theta} & \widetilde{\alpha}_{\theta}
     \end{pmatrix}(\vec{k}_1, \vec{k}_2)
     + O\left(\frac{1}{\sqrt{N}}\right) \, ,
 \end{align}
 and similarly for its hermitian conjugates $\mathbb{A}_{\theta}^{\dagger}$. On the other hand, $\mathbb{B}$ does not have linear terms. From the Bogoliubov transformation for bulk fields \eqref{eq:b_gff}, we can read off the corresponding transformation for large $N$ bi-local oscillators
 \begin{align}
     \alpha(\Vec{p}_1,\Vec{p}_2) =& \frac{2}{\cosh{2\theta_1}+\cosh{2\theta_2}}\left[\cosh{\theta_1}\cosh{\theta_2}\alpha_\theta(\Vec{p}_1,\Vec{p}_2)+\sinh{\theta_1}\sinh{\theta_2}\tilde{\alpha}_\theta^\dagger(\Vec{p}_1,\Vec{p}_2)\right] \, ,\\
     \gamma(\Vec{p}_1,\Vec{p}_2) =& \frac{2}{-\cosh{2\theta_1}+\cosh{2\theta_2}}\left[\cosh{\theta_1}\sinh{\theta_2}\gamma_\theta(\Vec{p}_1,\Vec{p}_2)+\sinh{\theta_1}\cosh{\theta_2}\tilde{\gamma}_\theta^\dagger(\Vec{p}_1,\Vec{p}_2)\right] \, ,
 \end{align}
and similarly for $\tilde{\alpha}$ and $\tilde{\gamma}$.
They annihilate the zero temperature vacuum state $\ket{0}$ and obey the regular commutation relations
 \begin{align}
     [\alpha(\Vec{p}_1,\Vec{p}_2), \alpha^\dagger(\Vec{p}_3,\Vec{p}_4) ] =& \delta^d(\Vec{p}_1-\Vec{p}_3)\delta^d(\Vec{p}_2-\Vec{p}_4) \, , \\ 
     [\gamma(\Vec{p}_1,\Vec{p}_2), \gamma^\dagger(\Vec{p}_3,\Vec{p}_4) ] =& \delta^d(\Vec{p}_1-\Vec{p}_3)\delta^d(\Vec{p}_2-\Vec{p}_4) \, ,
\end{align}
and similarly for $\tilde{\alpha}$ and $\tilde{\gamma}$. As a byproduct, the bulk fields can be expressed in terms of large $N$ bi-local oscillators in the same way as \eqref{eq:gff}, except that the subscripts $\theta$ are dropped on both hand-sides of the equation.

\paragraph{$H_{+}$ transformations.}
Let us now study in detail the total Hamiltonian $H_{+}$. It induces a non-unitary transformation, under which some certain combinations of operators annihilate the TFD state. According to \cref{eq:Hplus_transformation}, we find
\begin{equation} \label{eq:collective_Hplus_eg}
    \ee^{- \beta H_+ / 4} \big[ 
        a^{i}(\vec{k}_1) a^{i}(\vec{k}_2) - a^{i}(\vec{k}_1) \tilde{a}^{\dagger i}(\vec{k}_2)
    \big]
    \ee^{\beta H_+ / 4} \ket{0(\beta)} = 0 \, .
\end{equation}
These simple relations are the basics of the work \cite{Cottrell:2018ash} for building the TFD states. One can of course insert other oscillators $a^{\dagger i}$, $\tilde{a}^{i}$ and $\tilde{a}^{\dagger i}$ to make transformations of other collective fields. These transformations are easily computed in free theory, but not in interacting theories, in which case $H_+$ is quite involved. Nevertheless it is in principle calculable. To study the large $N$ limit, two procedures are involved: (1) $H_+$ transformation, and (2) the $1/N$ expansion in collective representation. The process (1) $\rightarrow$ (2) is well defined and one can in principle perform it. On the other hand, as we will see below, the problem is much harder if we perform (2) $\rightarrow$ (1), which resembles the translational issues in the soliton case. In the latter procedure, the $1/N$ expansion loses its power since we will need to taking the whole series of the $1/N$ expansion into account. This represents similar issues as in the soliton case.

\paragraph{}
The presence of this order $\sqrt{N}$ operator (i.e., $\sqrt{N} H_{+,\col}^{(1)}$) reveals that thermal backgrounds are not saddle point solutions of $H_{+, \col}$. Indeed, $H_{+, \col}$ has only one stationary point, which is also the minimum: the ground state. It meanwhile causes severe issues when applying to transformations of operators: it gives shifts of order $\sqrt{N}$ which is not defined in the large $N$ limit. As illustrated above, $W \equiv H_{+,\col}^{(1)}$ commutes with the thermofield Hamiltonian $\hat{H}$, so it becomes central in the large $N$ limit, as discussed by Witten \cite{Witten:2021unn}. Let us now take $\mathcal{O}$ as a collective operator which under the transformation of $H_{+}$ annihilates the TFD state, e.g. $\mathcal{O} = a^{i} a^{i} - a^{i} \tilde{a}^{\dagger i}$ as in \cref{eq:collective_Hplus_eg}. In the collective representation $\mathcal{O}_{\col} = \mathcal{O}$ can be expanded in $1/N$:
\begin{equation}
    \mathcal{O}_{\col} = \sum_{n=0}^{\infty} N^{1 - \frac{n}{2}} \mathcal{O}_{\col}^{(n)} \, .    
\end{equation}
However, to compute its $H_{+}$ transformation
\begin{equation}
    \ee^{- \frac{\beta}{4} H_{\col,+}} \mathcal{O}_{\col} \ee^{\frac{\beta}{4} H_{\col, +}} \, ,
\end{equation}
one has to do an infinite computation. This is because terms of order $\sqrt{N}$ and order $1/\sqrt{N}$ get mixed, so the evaluation can be done by expanding:
\begin{align} 
    \ee^{- \sqrt{N} \beta W / 4} \mathcal{O}_{\col} \ee^{\sqrt{N} \beta W / 4}
    & = \sum_{n=0}^{\infty} \sum_{m=0}^{\infty} N^{1 - \frac{n}{2} + \frac{m}{2}} 
    \frac{1}{m!} \left(-\frac{\beta}{4}\right)^{m} \ad_{W}^{m}(\mathcal{O}_{\col}^{(n)}) 
    \nonumber \\
    & = \sum_{r=0}^{\infty} N^{1 - \frac{r}{2}} \mathcal{O}_{\col}^{(r)} \, ,
\end{align}
with $\operatorname{ad}_{W}(\mathcal{O}) = [W, \mathcal{O}]$ denoting the adjoint transformation, and
\begin{equation}
    \mathcal{O}_{\col}^{(r)} = \sum_{n=r}^{\infty} \frac{1}{(n-r)!} \left(-\frac{\beta}{4}\right)^{n-r} \ad_{W}^{(n-r)}(\mathcal{O}_{\col}^{(n)}) \, . 
\end{equation}
Here to make simplification, we only expand $H_{+}$ to its order $\sqrt{N}$ term. The transformation thus reorganizes the all higher order terms $1/N$ expansion series into lower order ones. This is a universal property that does not only arise in the O$(N)$ vector model, but in TFD states for \emph{all} large $N$ theories. This is illustrated in the diagram below: the process (1) $\rightarrow$ (2) is calculable and gives a systematic $1/N$ expansion of the $H_{+}$ transformation of $\mathcal{O}$. The other process (2) $\rightarrow$ (1) is much more involved, and one needs to take the whole $1/N$ expansion series of $\mathcal{O}$ into account. 

\begin{center}
    \begin{tikzpicture}
        \node (A) {$\mathcal{O}$, $H_{+}$};
        \node (B) [right=5cm of A] {$\mathcal{O}_{\col}[\boldsymbol{\pi}, \boldsymbol{\eta}]$, $H_{+, \col}[\boldsymbol{\pi}, \boldsymbol{\eta}]$};
        \node (C) [below=3cm of A] {$\ee^{-\frac{\beta}{4}H_{+}} \mathcal{O} \ee^{\frac{\beta}{4}H_{+}}$};
        \node (D) [below=3cm of B] {$\mathcal{O}_{\col}^{(n)}$ transformed};

        \draw[->] (A.east) -- (B.west) node [black, midway, above] {(2) $1/N$};
        \draw[->] (C.east) -- (D.west) node [black, midway, below] {(2) $1/N$};
        \draw[->] (A.south) -- (C.north) node [midway, left] {(1) $H_{+}$};
        \draw[->] (B.south) -- (D.north) node [midway, right] {(1) $\sqrt{N} W$};
    \end{tikzpicture}
\end{center}

\section{Constraint of Thermofield Double State}
\label{sec:cst_tfd}

To recover the central part wave functional, using the bi-local operators
\begin{equation}
    C(\vec{k}_1, \vec{k}_2) = a^{i}(\vec k_1) \tilde{a}^i(\vec k_2) \, , \qquad
    C^{\dagger}(\vec{k}_1, \vec{k}_2) = a^{i \, \dagger}(\vec k_1) \tilde{a}^{i \, \dagger}(\vec k_2) \, \,
\end{equation}
We can write the $\hat{\mathcal G}(\vec k)$ operator as
\begin{equation}
    \hat{\mathcal{G}}(\vec k) = \ii \left[C^\dagger(\vec k, \vec k) - C(\vec k, \vec k)\right] \, .
\end{equation}
Recalled that the vacuum state is annihilated by $C$:
\begin{equation} \label{eq:C_vacuum}
    C(\vec k, \vec k) \ket{0} = 0 \, ,
\end{equation}
we can write \cref{eq:C_vacuum} as
\begin{equation}
    C^\dagger(\vec k, \vec k) + C(\vec k, \vec k) + \ii \hat{\mathcal G}(\vec k) \ket{0} = 0 \, .
\end{equation}
Thus we have
\begin{equation}
    C^{\dagger}_{\theta}(\vec k, \vec k) + C_{\theta}(\vec k, \vec k) + \ii \hat{\mathcal G}(\vec k) \ket{0(\beta)} = 0 \, .
\end{equation}
We note that the first two terms can be written in terms of the collective coordinate
\begin{equation}
    C^{\dagger}_{\theta}(\vec k, \vec k) + C_{\theta}(\vec k, \vec k) = 8 N \omega(\vec{k}) \hat{q}(\vec k) + \mathcal O(1) \, .
\end{equation}
Using the constraint we can also write the last term $\ii \hat{\mathcal G}(\vec k)$ as $\ii \hat{g}(\vec k)$. Thus we have
\begin{equation}
    \hat{q}(\vec k) + \frac{\ii}{8 \omega(\vec{k}) N} \hat{g}(\vec k) \ket{0(\beta)} = 0 \, .
\end{equation}
Note that $\hat{g}$ is of order $\mathcal{O}(\sqrt{N})$, and thus $\hat{q}$ is of order $\mathcal{O}(N^{-1/2})$.

\bibliographystyle{jhep}
\bibliography{reference}

\end{document}